\documentclass[11pt]{article}

\usepackage{./template/arxiv}

\usepackage{algorithm} 
\usepackage{algpseudocode} 
\usepackage{amsmath}
\usepackage{amsfonts}
\usepackage{amssymb}
\usepackage[english]{babel}
\usepackage{caption}
\usepackage{doi}
\usepackage{enumitem}
\usepackage{float}
\usepackage{graphicx}
\usepackage{hyperref} 
\usepackage[capitalise]{cleveref}
\usepackage{mathptmx}
\usepackage{microtype}
\usepackage{subfig}
\usepackage{tabularx}
\usepackage[utf8]{inputenc}

\usepackage{booktabs}
\usepackage{siunitx}
\usepackage{xspace}
\usepackage{cite}

\usepackage{./input/luh-colors}
\usepackage{./input/ao-math-std}
\usepackage{./input/ao-math-graphs}
\usepackage{./input/ao-math-fields}

\newcommand\orcauth[2]{\href{https://orcid.org/#1}
  {\includegraphics[height=0.7em]{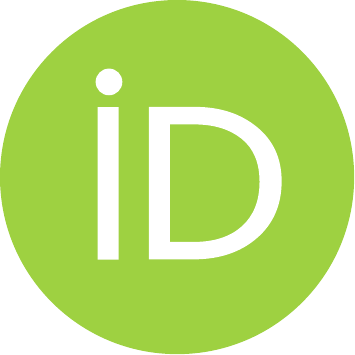}\enspace#2}}
\newcommand\con{\tsb{con}}
\newcommand\geo{\tsb{geo}}
\newcommand\perf{\tsb{perf}}
\newcommand\term{\tsb{term}}
\newcommand\CCO{\tsp{in}}
\colorlet{imgcol}{LUH-green}
\newcommand\image[2][]{\bgroup\fboxsep0pt\fboxrule0.2pt
  \protect\fcolorbox{imgcol}{imgcol!10}{\protect\includegraphics[#1]{#2}}\egroup}

\title{Rigorous mathematical optimization of synthetic hepatic vascular trees}
\date{}

\author{
    \orcauth{0000-0003-0276-6029}{Etienne Jessen}\\
	Institute of Mechanics and Computational Mechanics\\
	Leibniz Universität Hannover\\
	30167 Hannover, Germany\\
	\texttt{etienne.jessen@ibnm.uni-hannover.de} \\
\And
    \orcauth{0000-0002-6343-9809}{Marc C. Steinbach} \\
	Institute of Applied Mathematics\\
	Leibniz Universität Hannover\\
	30167 Hannover, Germany\\
	\texttt{mcs@ifam.uni-hannover.de} \\
\And
    \orcauth{0000-0003-1962-238X}{Charlotte Debbaut} \\
	IBiTech – Biommeda \\
	Ghent University\\
	Ghent, Belgium \\
	\texttt{Charlotte.Debbaut@UGent.be} \\
\And
    \orcauth{0000-0002-9068-6311}{Dominik Schillinger} \\
	Institute of Mechanics and Computational Mechanics\\
	Leibniz Universität Hannover\\
	30167 Hannover, Germany\\
	\texttt{schillinger@ibnm.uni-hannover.de} \\
}

\colorlet{imgcol}{white}
\begin{document}
\maketitle

\begin{abstract}
  In this paper,
  we introduce a new framework for generating synthetic vascular trees,
  based on rigorous model-based mathematical optimization.
  Our main contribution is the reformulation of finding the optimal global tree geometry
  into a nonlinear optimization problem (NLP).
  This rigorous mathematical formulation accommodates efficient solution algorithms
  such as the interior point method
  and allows us to easily change boundary conditions and constraints applied to the tree.
 Moreover, it creates trifurcations in addition to bifurcations.
  A second contribution is the addition of an optimization stage for the tree topology.
  Here, we combine constrained constructive optimization (CCO)
  with a heuristic approach to search among possible tree topologies.
  We combine the NLP formulation and the topology optimization into a single algorithmic approach.
  Finally, we attempt the validation of our new model-based optimization framework
  using a detailed corrosion cast of a human liver,
  which allows a quantitative comparison of the synthetic tree structure
  to the tree structure determined experimentally down to the fifth generation.
  The results show that our new framework is capable of
  generating asymmetric synthetic trees that match
  the available physiological corrosion cast data better than trees
  generated by the standard CCO approach.
\end{abstract}


\newpage
\section{Introduction}
\begin{figure}[h]
  \centering
  \image[trim=0 5 0 5,clip=true,width=0.86\textwidth]
  {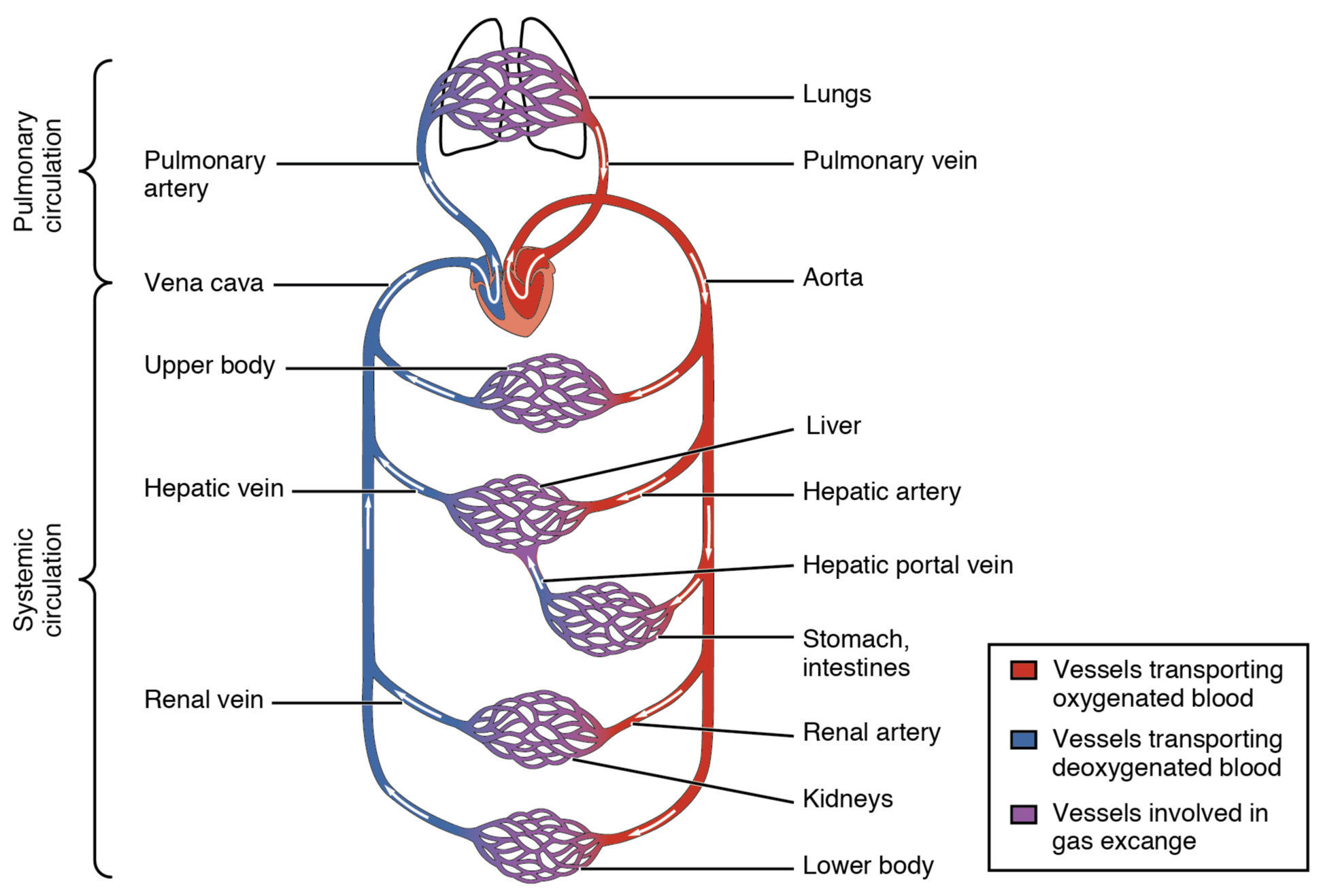}
  \caption{Schematic overview of the cardiovascular system \cite{openstax}}
  \label{fig:cardiovascular_system}
\end{figure}

The cardiovascular system of the human body supplies the cells with vital nutrients
by permitting blood to circulate throughout the body \cite{noordergraaf2012circulatory}.
The heart pumps the blood through vessels,
categorized into arteries (transporting blood away from the heart)
and veins (transporting blood towards the heart).
The cardiovascular system is further divided into the pulmonary circulation
and the systemic circulation.
In the pulmonary circulation, deoxygenated blood is carried from the heart to the lungs,
and oxygenated blood returns to the heart.
In contrast, the systemic circulation carries oxygenated blood
from the heart to the rest of the body, reaching the other organs.
The blood enters these organs through different branches of the aorta, where arteries distribute it.
The arteries split into
smaller and smaller arteries until they reach the arterioles, which are the last arterial branches prior to entering the microcirculation.
After the blood is distributed at the microcirculatory level and interacts with the organ's cells,
the capillaries merge to bring the deoxygenated blood back through the venules, which merge into veins.
Finally, the blood leaves the systemic circulation
through either the superior or inferior vena cava back to the heart.
The complete cardiovascular system is schematically shown in \cref{fig:cardiovascular_system}.

\pagebreak[1]
Formally, the systemic circulation can be divided into two functional parts: macrocirculation and microcirculation.
In the microcirculation, nutrients and oxygen diffuse towards the organ's cells.
Here, the main functions of the different organs are carried out, e.g.,
synthesizing proteins and detoxification in the liver.
In contrast, macrocirculation mainly distributes oxygenated blood evenly throughout the organs
and then recollects the deoxygenated blood.
The task of distributing and collecting blood leads to specific branching patterns inside the organs.
These sets of branches, at least one for arteries and one for veins inside each organ, are called vascular trees.
The general structure of a vascular tree mainly depends on the organ supplied,
with the main factors being the organ's shape, the amount of blood supply, and the microcirculation structure.
Furthermore, a distinction between solid organs (such as the
liver) and hollow organs (such as the stomach) must be made.

\pagebreak[1]
In general, vascular trees are patient-specific,
and clinicians cannot derive them from statistical measures alone.
Having detailed patient-specific data on vascular trees
would be essential to help further improve many clinical treatment strategies,
for example determining suitable cut patterns in liver resection
or optimizing targeted chemo-therapy for cancer patients.
An essential tool for obtaining patient-specific data on vascular trees in vivo
is noninvasive medical imaging such as CT or MRI.
Their maximum resolution for in-vivo imaging, however,
even with the advances made in the last decade, is still limited.
Therefore, for understanding vascular trees down to the arterioles and venules,
ex-vivo methods must be used which are often time-consuming, require
specialized equipment, and are therefore expensive.
Examples are cryomicrotomes in human hearts \cite{goyal2012model} or corrosion casting of the liver \cite{debbaut2014analyzing}.

An alternative approach is based on the synthetic generation of vascular trees with the help of a computer.
Starting from available low-resolution patient-specific imaging data,
synthetic vascular trees can potentially fill in the missing data
to obtain a high-resolution model-based representation of the hierarchical vascular system.
These synthetic vascular trees are based on optimality principles whose goal is to minimize the metabolic cost
\cite{zamir2002physics,murray1926physiological,cassot2010branching}.
The assumption is that the individual branchings, defining the structure of the tree on the macroscale,
form under these principles.
Most existing methods generate vascular trees based on these optimization principles
and assume that flow is distributed evenly into a pre-defined perfusion volume.
Such synthetic trees can be generated to any pre-arteriolar
refinement level.
A number of different methods exist that differ in terms of the optimization algorithms
and the constraints for guiding the optimization.

The most well-known approach for generating vascular trees is the Constrained Constructive Optimization (CCO) method,
first proposed by Schreiner et al.\ \cite{schreiner1993computer} and later extended to three-dimensional
non-convex domains \cite{karch1999three}.
It is based on modeling blood flow using Poiseuille's law
and utilizing Murray's law \cite{murray1926physiological2} for optimizing bifurcations.
The underlying algorithm starts from an initial vascular structure and iteratively adds new segments
while optimizing the local tree structure (topology and geometry) at each bifurcation.
Since CCO plays a central role in the topology optimization of our framework, we review the method in more detail later.
CCO can reproduce a qualitatively reasonable distribution of segments, but fails to capture the asymmetric branching patterns
that characterize most real vascular trees.
Several adaptations to CCO have been proposed that attempt to remove this limitation,
for example using new constraints or new intermediate processing steps
for generating organ-specific vascular systems \cite{talou2021adaptive,jaquet2018generation,schwen2012analysis}.
Moreover, due to the sampling of new segments
the results of CCO-generated trees are largely dependent on random seeds \cite{schreiner2003heterogeneous}.
In Georg et al.\ \cite{hahn2005fractal}, an alternative method known as Global Constructive Optimization (GCO)
was introduced.
It starts by defining random points inside the perfusion volume.
These points are kept fixed throughout the optimization and are the leaf nodes of the resulting vascular tree.
The goal is now to construct the topology and positions of the internal nodes of the vascular tree.
Optimization is driven by successively connecting all leaf nodes to existing internal nodes
(starting with only the root node) and then using splitting and pruning steps to create new internal nodes.
Thus the method iterates through suboptimal global structures
until the tree reaches a suitable level of refinement.
Organ-specific methods for generating vascular trees have also been introduced, e.g.,
for the stomach \cite{talou2021adaptive}
and the liver \cite{schwen2012analysis, schwen2015algorithmically, rohan2018modeling}.

\pagebreak[1]
A recent method, proposed by Keelan et al.\ \cite{keelan2016simulated},
is based on the assumption
that the limitation in the results of CCO and its variants
are caused by the fact that only optima of the local tree structure are explored.
Instead of adding intermediate or postprocessing steps to CCO,
a new approach based on simulated annealing (SA) was introduced
to search for the optimum of the global tree structure.
Like GCO, this approach generates the leaf nodes beforehand.
The optimization step consists in adjusting the topology
and geometry of the vascular tree iteratively.
It was claimed that the approach will converge
against the global minimum if the number of iterations goes to infinity.
Results also show a visual convergence of trees
with different initial structures
to very similar global structures after optimization,
a feature no other introduced method was able to reproduce.
However, as simulated annealing is used
for both topology and geometry optimization,
the algorithm is extremely costly for decently sized three-dimensional vascular trees
and global convergence cannot be guaranteed.

\pagebreak[1]
In this paper, we introduce a new framework for generating synthetic vascular trees,
which rigorously mitigates the limitations of the CCO approach, achieving
results similar to the SA based method
but at a significantly lower computational cost.
We start by casting the problem of finding the optimal global tree geometry
into a nonlinear optimization problem (NLP).
We then specialize the global model for optimizing the local geometry
of a single new branching.
This rigorous mathematical formulation accommodates efficient solution algorithms
and makes changes in boundary conditions and constraints trivial.
The framework also includes a discrete optimization step for finding a near-optimal tree topology.
To this end, it combines CCO with a heuristic subtree swapping step
motivated by the SA approach \cite{keelan2016simulated}.
We combine the geometry and topology optimization steps into a single algorithmic approach.
Unlike the standard CCO approach and its variants,
we reduce the resulting volume of the tree significantly and limit the influence of random samples
on the final global tree structure.
Based on the formal separation of topology and geometry optimization,
the efficiency of the algorithm is significantly improved compared to the SA approach.
The new framework allows us to generate a synthetic tree inside a non-convex organ up to the pre-arteriolar level,
where the microcirculation starts and the tree
transmutes into a meshed network of micro-vessels.

\section{Methods}

\subsection{Model assumptions}

We model the vascular tree
as a branching network $\Tree = (\Nodes, \Arcs)$,
consisting of nodes $u \in \Nodes$ and segments $a \in \Arcs$.
The segments are assumed to be rigid and straight cylindrical tubes,
and each segment $a = uv$ is defined by its radius $r_a$
and the geometric locations of its proximal node $x_u$ and distal node $x_v$,
yielding the length $\ell_a = \norm{x_u - x_v}$.
The goal is to generate the vascular tree
inside a given (non-convex) perfusion volume $\Omega \subset \R^3$,
while homogeneously distributing all terminal nodes
(\emph{leaves}) $v \in \Leaves$.
The network is perfused at steady state by blood,
starting at the feeding artery (\emph{root segment})
down to the leaves at the \emph{terminal segments}.
In a real vascular system, the tree transmutes into
an arcade network of micro-vessels \cite{peeters2017multilevel}
(mathematically a general meshed graph with cycles)
when reaching the arteriolar level
(radii in the range of \SIrange{0.02}{0.1}{mm}).
As such, the pre-arteriolar level marks a conceptual cut-point
of this model since the underlying assumptions
are no longer justified \cite{schreiner1993computer}.
To simplify the model, blood is assumed to be
an incompressible, homogeneous Newtonian fluid.
Further assuming laminar flow, we can express the hydrodynamic resistance $R_a$ of segment $a$ by Poiseuille's law as
\begin{equation}
  R_a = \frac{8 \eta}{\pi} \frac{\ell_a}{r_a^4} \quad\forall a \in \Arcs,
\end{equation}
where $\eta$ denotes the dynamic viscosity of blood
which is assumed constant with $\eta = \SI{3.6}{cP}$.
We note, however, that the typical radius of the smallest arteries
in the pre-arteriolar level
is in the range \SIrange{0.1}{0.2}{mm},
and the so-called Fåhræus--Lindqvist effect \cite{pries1994resistance}
should be taken into account for these vessels with
\begin{align}
  \eta(r_a)
  &= 1.125 \bigl( \kappa + \kappa^2 \bigl[
    6 \exp(-170r_a / \si{mm}) - 2.44 \exp(-8.09 (r_a / \si{mm})^{0.64}) + 2.2
    \bigr] \bigr), \\
  \kappa &= \frac{r_a^2}{(r_a - \SI{0.00055}{mm})^2}.
\end{align}
This effect describes the change of the blood viscosity based on the vessel diameter and,
in particular, the decrease of viscosity as the vessel diameter decreases.
This stems from the fact that in smaller vessels the blood cells tend to be in the center,
forcing plasma towards the walls, which decreases the peripheral friction.
The pressure drop $\Delta p_a$ over segment $a$ can now be computed by
\begin{equation}
  \Delta p_a = R_a Q_a \quad\forall a \in \Arcs,
\end{equation}
where $Q_a$ is the volumetric blood flow through segment $a$.
At individual branchings, the relationship between a parent segment
and its daughter segments obeys the power law
\begin{equation}
  r_{uv}^\gamma = \sum_{vw \in \Arcs} r_{vw}^\gamma
  \quad\forall v \in \Nodes \setminus \Leaves,
\end{equation}
where $\gamma$ is the \emph{branching exponent}.
It has the value $3.0$ in Murray's law \cite{murray1926physiological2}, which is
shown to yield a balance between minimizing metabolic cost
of maintaining blood and power loss
for moving blood \cite{horsfield1989diameters}.
In the literature, $\gamma$ values from $2.0$ to $3.0$
are generally considered valid for vascular trees
\cite{vanbavel1992branching,van20133d,zhou1999design,kurz1997modelling,
  godde2001structural},
with, e.g., $\gamma = 2.55$ minimizing pulsative flow \cite{kurz1997modelling}
and $\gamma = 2.7$ minimizing vascular wall material
\cite{godde2001structural}.
As noted in Schwen et al.\ \cite{schwen2012analysis},
a constant value $\gamma$ might not be very realistic and $\gamma$
should be considered dependent on the branching generation in the future.

In addition to the model assumptions, a set of physiological constraints
are needed to construct the vascular tree.
As suggested in Schreiner et al.\ \cite{schreiner1993computer},
we assume that the tree minimizes the metabolic cost of maintaining blood inside the tree,
which is proportional to the tree's volume,
\begin{equation}
  f_\Tree = \sum_{a\in\Arcs} \pi \ell_a r_a^2.
\end{equation}
We further constrain the tree to have equal pressure $p\term$ at all
terminal nodes, which are the entry points into the microcirculatory network.
Since the tree induces a given total perfusion $Q\perf$ (at the root node)
across an overall pressure drop $\Delta p = p\perf - p\term$, this constraint leads
to equal outflow at each terminal node.

\subsection{Constrained constructive optimization (CCO)}
\label{sec:CCO}

Before we introduce our framework,
we first describe CCO in more detail and illustrate key properties of its results
via a representative benchmark example.
We note that for visualizing trees, we employ the software POV-Ray \cite{povray},
where we represent branching points as spheres and segments as cylinders.

\subsubsection{Algorithmic background and key properties}

\begin{figure}
  \centering
  \subfloat[Sampling of new terminal point]
  {\image[trim=0 0 5 10,clip=true,width=0.365\textwidth]{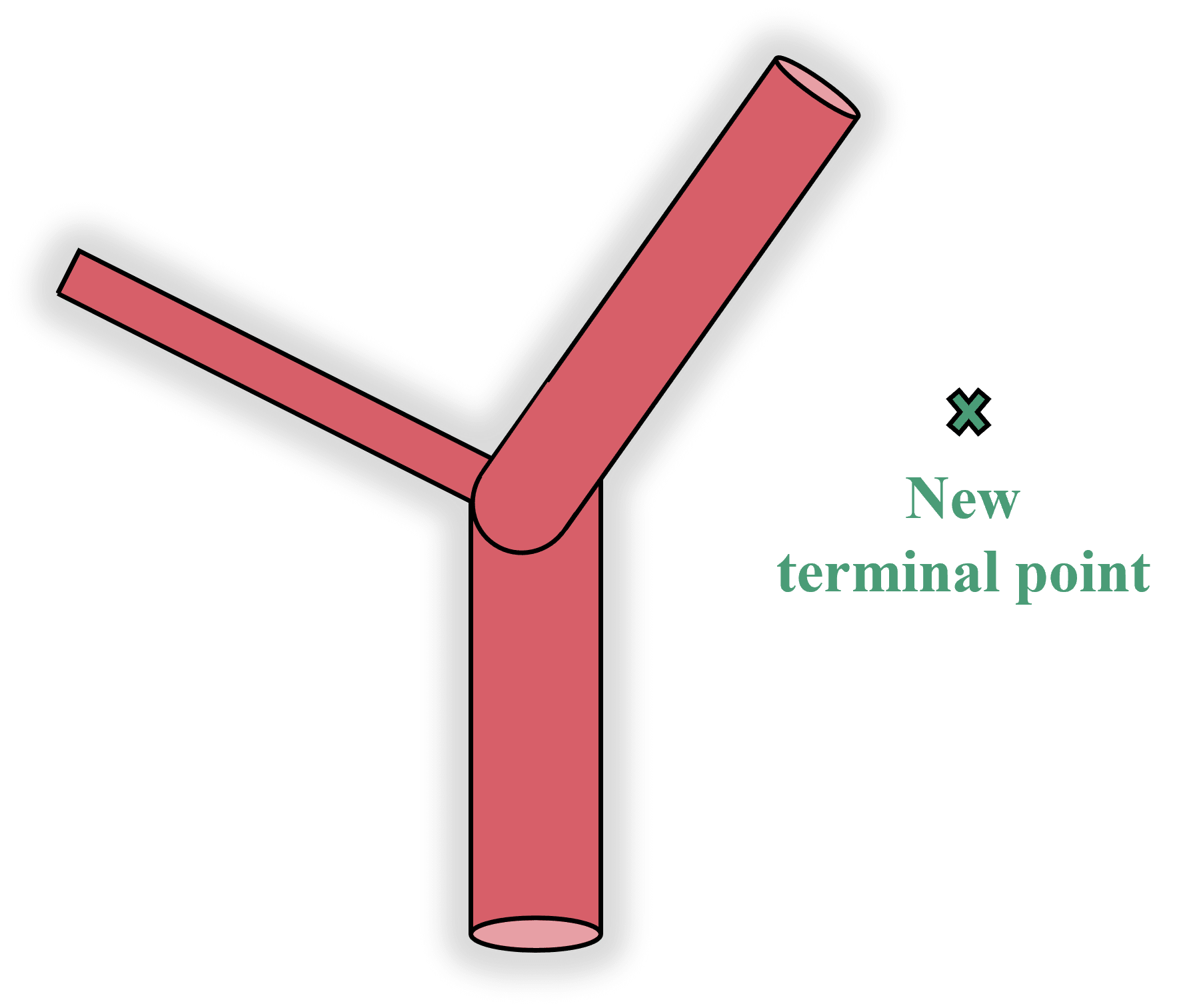}}\hfill
  \subfloat[Connection to existing segment]
  {\image[width=0.31\textwidth]{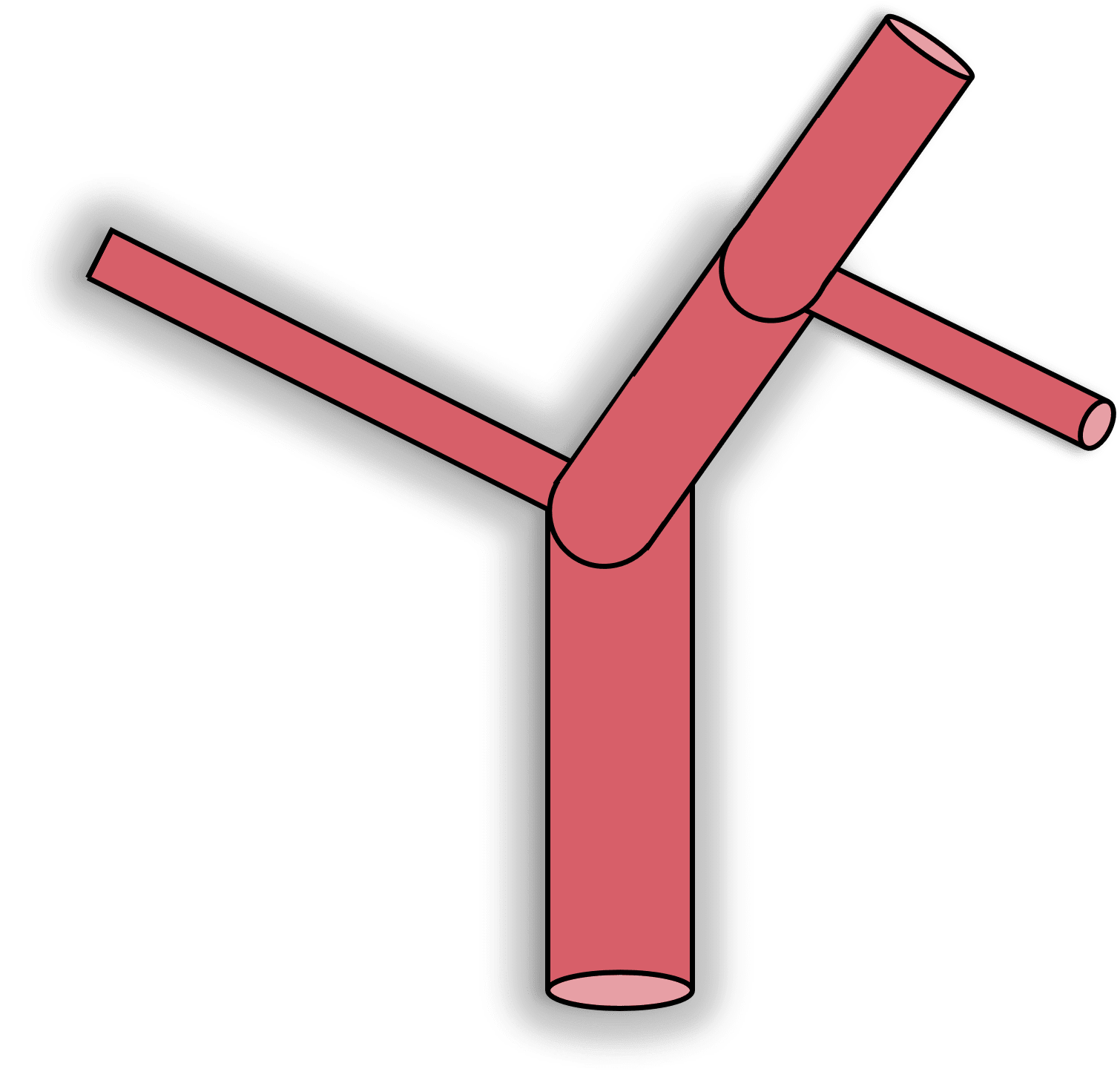}}\hfill
  \subfloat[Optimization of new bifurcation]
  {\image[width=0.31\textwidth]{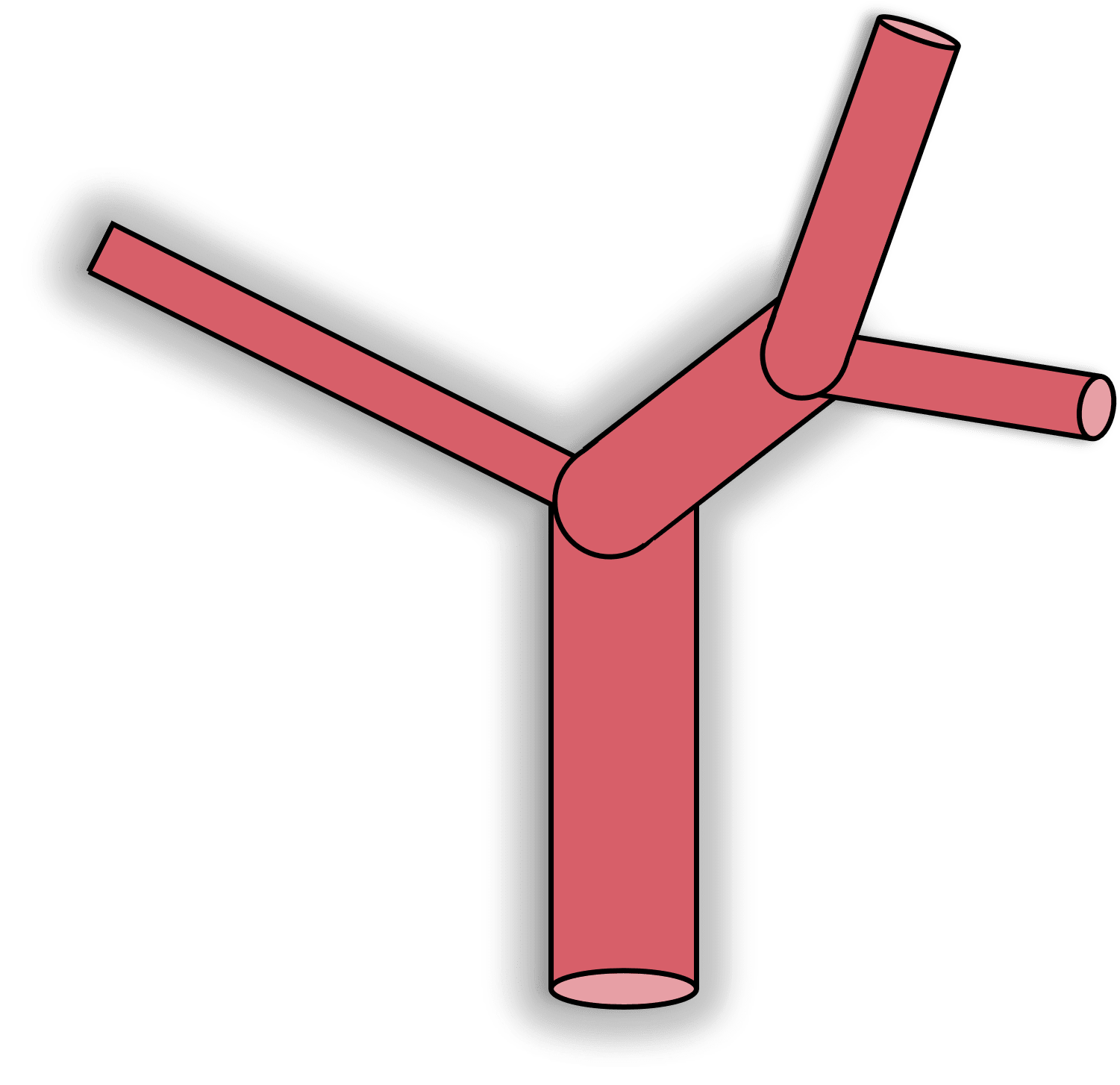}}\\
  \caption{Schematic overview of CCO's growth algorithm, showing the three main steps.
    Steps (b) and (c) are repeated for all
    neighboring segments of the new terminal point.}
  \label{fig:CCO_schematic}
\end{figure}

CCO can generate a vascular tree
under the assumptions and boundary conditions described above.
The main idea behind CCO is to grow the vascular tree incrementally
by adding new segments one by one.
Each addition consists of three steps.
Step~1 is to sample a new terminal point $x\term$ uniformly
inside the perfusion volume.
The distance of the sampled point to each existing segment
must be larger than a pre-defined threshold.
This threshold ensures that the new terminal point
is compatible with the current tree geometry
and leads to a uniform distribution of all terminal points
inside the perfusion volume.
The distance between the sampled point and a segment is computed
by evaluating the orthogonal projection onto the convex line segment.
The required threshold is lowered
with each iteration to accommodate the growing number of segments inside the perfusion volume.
After a new terminal point is found,
it is connected to an existing segment in step~2,
leading to a new bifurcation.
In step~3, the location of this newly created bifurcation
is optimized for the lowest total tree volume.
Steps 2 and~3 are repeated only for the $N\con$ closest segments of $x\term$,
and the connection with the lowest total volume is chosen as permanent.
The number $N\con$ of different connections tested will be investigated later.
The entire approach is visualized in \cref{fig:CCO_schematic}.
The search for the best connection is an optimization of the local topology,
while the search for the best location of the bifurcation point
is an optimization of the local geometry of the tree.

The growth algorithm for a tree can either run
until a prescribed number of segments are connected
or until the radius of new terminal segments is below a certain threshold
(usually the minimum radius of the pre-arteriolar level,
$r\tsb{min} = \SI{0.1}{mm}$).
The main computational burden of CCO is the geometry optimization
that follows the introduction of the new bifurcation in each iteration.
After a new terminal point is connected to the tree,
the constraints and boundary conditions (e.g., equal terminal outflow)
do not hold any longer.
The hydrodynamic resistance of each segment
on the path from the new bifurcation to the root
needs to be rescaled to account for this newly created segment,
subsequently inducing a rescaling of the root radius.
Therefore, all radii need to be recomputed.
This rescaling of the tree is a recursive computation
starting from the new terminal segment,
which is described in detail in Karch et al.\ \cite{karch1999three}.
Each time the position of a bifurcation is changed,
the tree needs to be rescaled in such a manner.

Synthetic trees generated by the CCO approach
show good visual agreement with morphological data
and have comparable mean radii over all generations.
However, one of the most significant drawbacks of CCO
is the inability to generate trees with asymmetric bifurcation ratios.
In vascular systems, blood is transported
over long distances inside bigger arteries,
while only being in small arteries for a short distance.
This leads to direct connections between small arteries and large trunks
and to small bifurcation ratios.
Only when approaching the smallest arteries,
a shift to larger bifurcation ratios can be observed.
In contrast to these specific structures,
CCO-generated trees tend to be more symmetric across all segments
with flow evenly splitting into both branch segments.
Many augmented versions of CCO were proposed to tackle this,
often introducing postprocessing steps and new constraints.

\subsubsection{Representative benchmark example}

To summarize important characteristics of CCO-generated vascular trees
and to establish a consistent way of quantifying them,
we apply standard CCO to the benchmark problem
introduced in Karch et al.\ \cite{karch2000staged}.
The perfusion volume is a shallow rectangular box,
and the root node is located at one of the corners.
The model parameters are summarized in \cref{tab:benchmark_parameters}.

\begin{table}
  \centering
  \caption{Model parameters
    of the benchmark problem due to Karch et al.\ \cite{karch2000staged}}
  \label{tab:benchmark_parameters}
  \begin{tabular}{*2{l@{\qquad}}l}
    \toprule
    Parameter & Meaning & Value\\
    \midrule
    $V\perf$ & perfusion volume & \SI{9}{cm} x \SI{7}{cm} x \SI{1.6}{cm} \\
    $p\perf$ & perfusion pressure & \SI{100}{mm Hg} \\
    $p\term$ & terminal pressure & \SI{60}{mm Hg} \\
    $N\term$ & number of terminal segments & 6,000 \\
    $Q\perf$ & perfusion flow (at root) & \SI{500}{ml/min} \\
    $\eta$   & blood viscosity & \SI{3.6}{cP} \\
    $\gamma$ & branching exponent & $2.55$ \\
    $N\con$  & maximum number of connection tested
             & $\set{2, 4, 8, 16, 32, 64, 128}$ \\
    \bottomrule
  \end{tabular}
\end{table}

\begin{figure}
  \centering
  \image[trim=0 15 0 10,clip=true,width=0.5\linewidth]
    {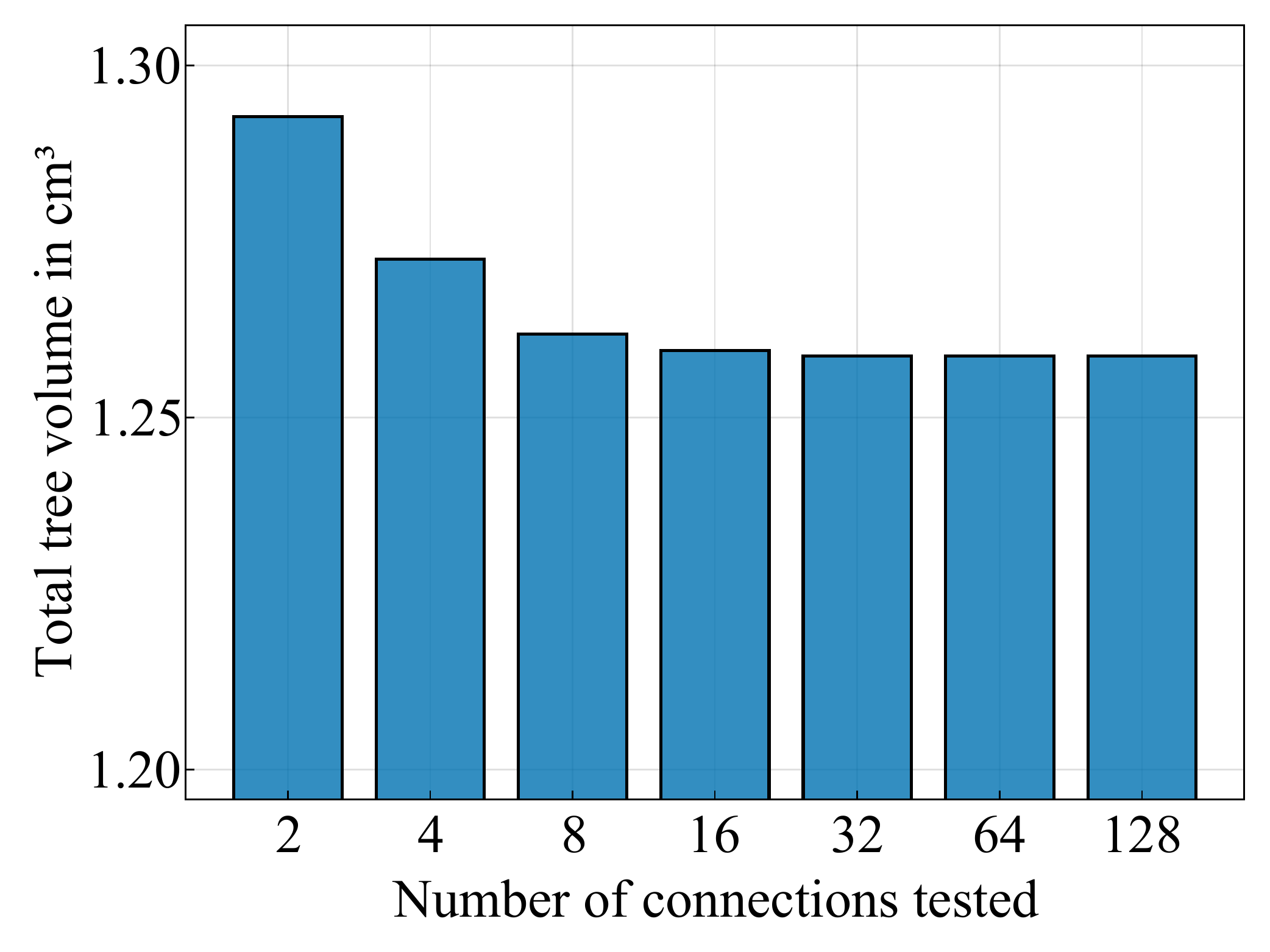}
  \caption{Total tree volume for different numbers
    of connection tests ($N\term = 6{,}000$)}
  \label{fig:CCO_benchmark_connection_test}
\end{figure}

As stated above,
CCO performs an optimization of the local tree structure.
The topology optimization consists of connecting
a newly sampled terminal node to different segments one after another.
Only neighboring segments are connected,
and a maximum number $N\con$ of connections
is tested to make the computation more efficient.
To determine an appropriate choice for $N\con$ in our example,
we generated seven trees with different values of $N\con$,
summarized in \cref{tab:benchmark_parameters}.
The total volumes of the resulting trees are compared
in \cref{fig:CCO_benchmark_connection_test}.
Our results suggest a value of $N\con = 32$,
as testing more connections
had no significant influence on the final tree volume
while increasing the overall computation time.
We note that we will also use the value $N\con = 32$ for all further computations throughout the paper,
including those in the context of our new framework that we will introduce below.

\begin{figure}
  \centering
  \subfloat[$N\term = 50$]{\image[trim=100 30 25 40,clip=true,
      width=0.33\textwidth]{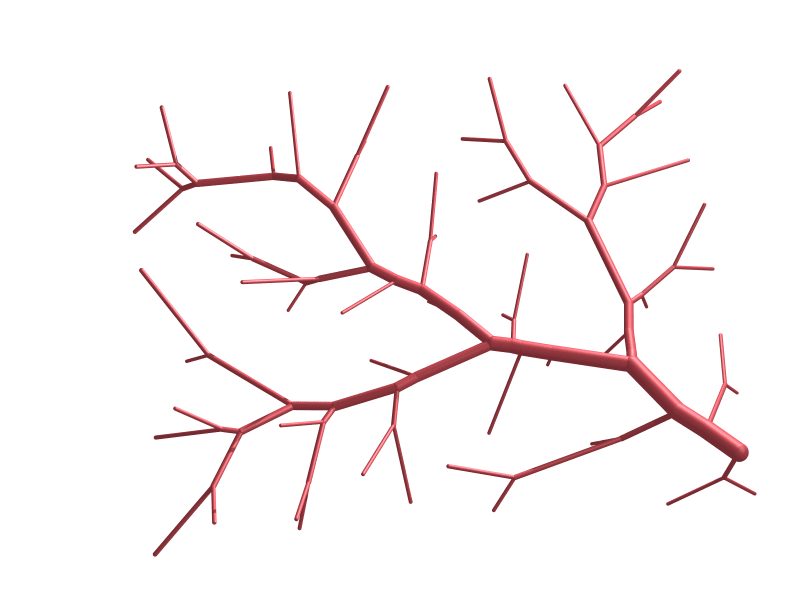}}\hfill
  \subfloat[$N\term = 500$]{\image[trim=100 30 25 40,clip=true,
      width=0.33\textwidth]{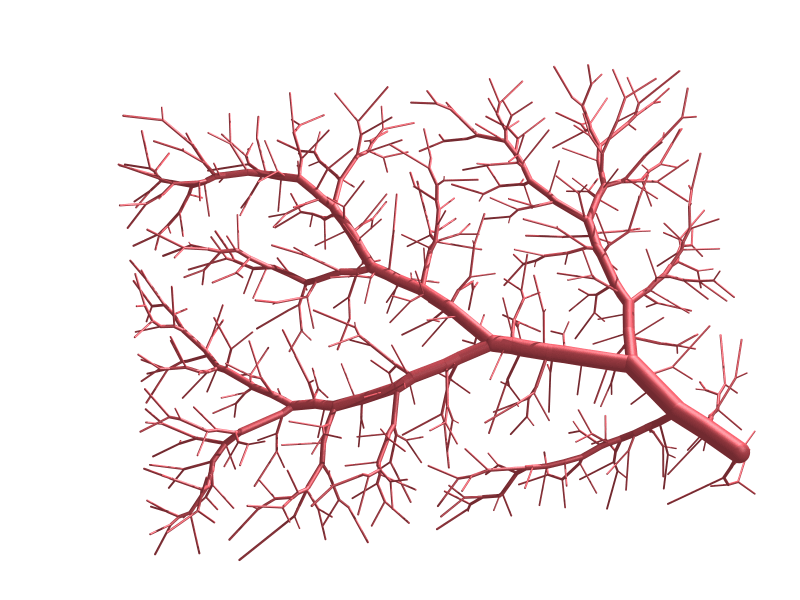}}\hfill
  \subfloat[$N\term = 6{,}000$]{\image[trim=100 30 25 40,clip=true,
      width=0.33\textwidth]{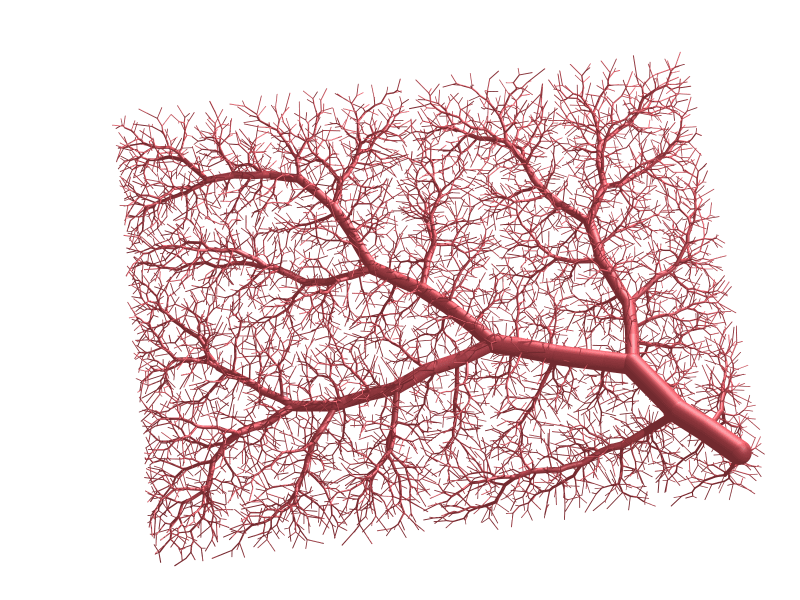}}\\
  \caption{Different stages of a synthetic tree during CCO-driven growth}
  \label{fig:CCO_benchmark_initial_sampling}
\end{figure}

Due to the iterative nature of the CCO approach,
segments that are generated early on
tend to define the overall hierarchy of the final tree.
This phenomenon is illustrated in \cref{fig:CCO_benchmark_initial_sampling}
for different numbers of terminal points.
We observe that after adding 50 terminal points only,
the core structure is nearly identical to that
of the final tree with 6,000 terminal points.
The reason is that CCO only changes the position
of one bifurcation in each iteration.
Therefore positions of old bifurcations are fixed,
and the corresponding segments do not change after initial generation.
All previously optimized bifurcations, however, are no longer optimal after the next iteration.
Furthermore, employing only subsequent disconnected geometry optimization steps
tends to favor symmetric bifurcations, even for segments that appear further down the tree hierarchy.

\begin{figure}
  \centering
  \subfloat[Seed 1]{\image[trim=100 30 25 40,clip=true,
      width=0.33\textwidth]{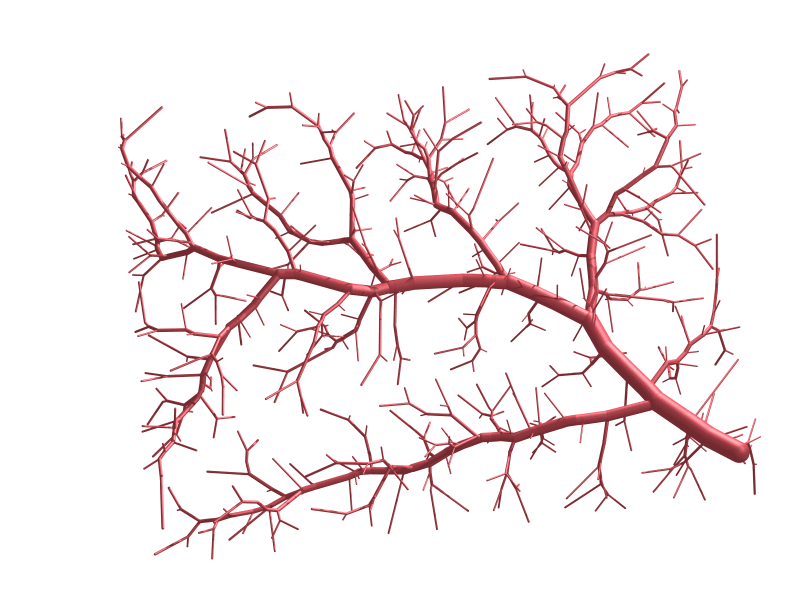}}\hfill
  \subfloat[Seed 2]{\image[trim=100 30 25 40,clip=true,
      width=0.33\textwidth]{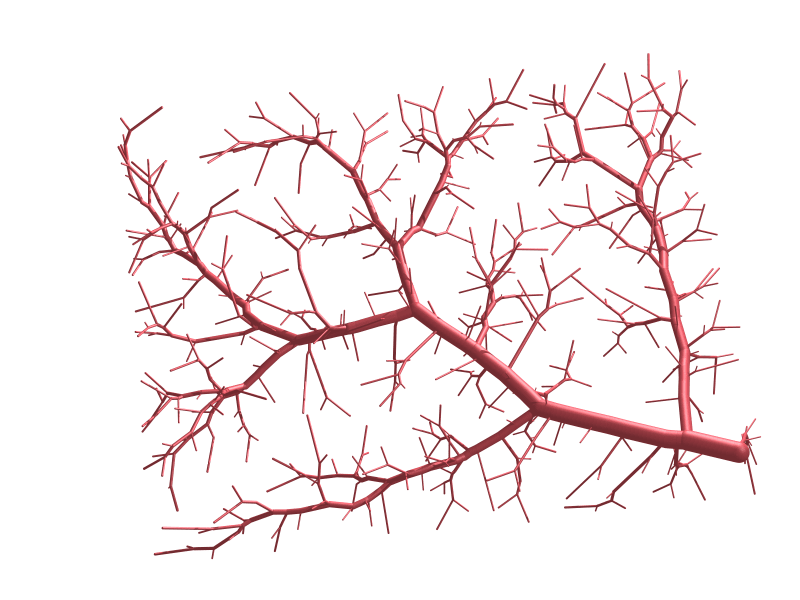}}\hfill
  \subfloat[Seed 3]{\image[trim=100 30 25 40,clip=true,
      width=0.33\textwidth]{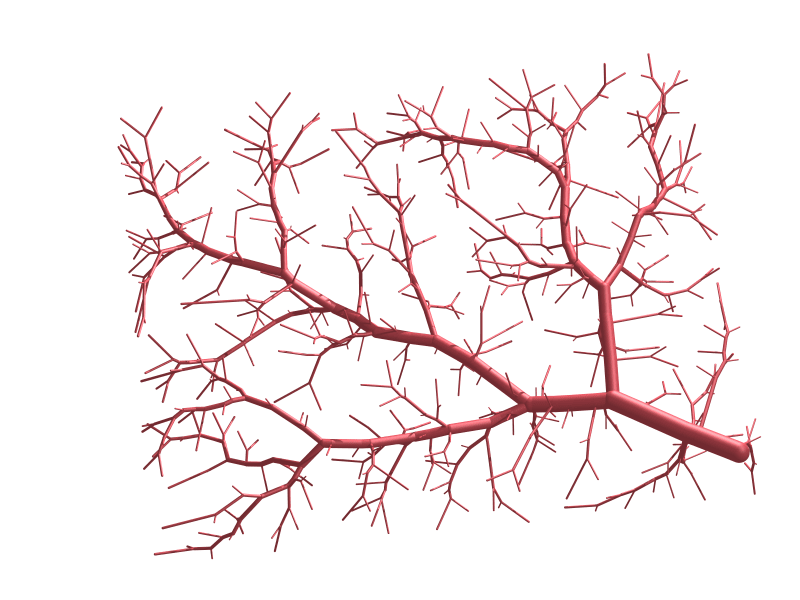}}\\
  \caption{Different random seeds
    during CCO-driven growth ($N\term = 500$)}
  \label{fig:CCO_benchmark_random_results}
\end{figure}

The bias of missing re-adjustment after adding new bifurcations is further amplified
by the bias of the specific random seed on the initial sampling
and the order in which samples are connected.
This bias is illustrated in \cref{fig:CCO_benchmark_random_results}, where we used the same terminal points for each tree but connected them in the order defined by their random seed. We can observe that three different random seeds lead to three very different tree structures.
Due to the dependence of the tree's topology on the sampled terminal points, only
qualitative comparisons are possible.
A quantitative comparison of the exact segment locations against a real vascular system is not possible
because results of the CCO method are not reproducible without pre-defining a fixed random seed.

\subsection{A new approach based on optimizing the global geometry}

The drawbacks of CCO are all due, at least partially,
to optimizing only the local tree structure at each bifurcation.
To mitigate this significant limitation and the associated problems,
we introduce a new framework for generating synthetic trees that optimizes their geometry and topology.
To this end, we formulate a nonlinear optimization problem (NLP)
to optimize the \emph{global} tree geometry, considering
\emph{all} branchings simultaneously.
We furthermore add a heuristic step for the optimization of the tree topology.
We cast these optimization steps into an algorithmic framework
that uses CCO as a tool to grow the tree in between these optimization steps.

\subsubsection{Geometry optimization}
We start with a CCO generated (near-optimal) tree
$\Tree = (\Nodes, \Arcs)$ whose continuous variables serve
as initial estimate of the global geometry.
We assume that we are given (for instance via medical imaging) the root subtree of depth $k$
with topology $\Tree_k = (\Nodes_k, \Arcs_k)$,
node locations $\_x_u$, $u \in \Nodes_k$,
as well as segment radii $\_r_a$ and lengths
$\_\ell_{uv} = \norm{\_x_u - \_x_v}$, $a = uv \in \Arcs_k$.
If $k = 0$, only the root location $\_x_0$ is provided.
Locations $\_x_u$ of all terminal nodes $u \in \Leaves$
are given by sampling their spatial distribution.
To circumvent the computationally expensive recursive computation
of the radii $r = (r_a)_{a \in \Arcs}$ as in Karch et al.\ \cite{karch1999three}
and similarly of the node pressures $p = (p_u)_{u \in \Nodes}$,
we include them together with the lengths $\ell = (\ell_a)_{a \in \Arcs}$
in the vector of optimization variables,
$y = (x, p, \ell, r)$, where $x = (x_u)_{u \in \Nodes}$.
We have physical lower bounds $\ell^-, r^-$ on $\ell_a, r_a$, respectively,
and we add artificial upper bounds $\ell^+, r^+$ for numerical efficiency.
Then $y$ has to be an element of the box
$Y = \R^{4 \card\Nodes} \x [\ell^-, \ell^+]^{\card\Arcs} \x [r^-, r^+]^{\card\Arcs}$
of dimension $4 \card\Nodes + 2 \card\Arcs = 6 \card\Nodes - 2$,
and our NLP reads:
\begin{align}
  \min_{y \in Y} \quad
  & \sum_{a \in \Arcs} \ell_a r_a^2 \\
  \stq
  \label{eq:nlp-fix-x}
  &0 = x_u - \_x_u, & u &\in \Nodes_k \cup \Leaves \\
  &0 = \ell_a - \_\ell_a, & a &\in \Arcs_k \\
  \label{eq:nlp-fix-r}
  &0 = r_a - \_r_a, & a &\in \Arcs_k \\
  \label{eq:nlp-length}
  &0 = \ell_{uv}^2 - \norm{x_u - x_v}^2, & uv &\in \Arcs \setminus \Arcs_k \\
  \label{eq:nlp-murray}
  &0 = r_{uv}^\gamma - {\textstyle\sum}_{vw \in \Arcs} r_{vw}^\gamma,
  &v &\in \Nodes \setminus (\Nodes_k \cup \Leaves) \\
  \label{eq:nlp-deltap}
  &0 = p_u - p_v - (8 \eta / \pi) Q_{uv} \ell_{uv} / r_{uv}^4 &uv &\in \Arcs \\
  \label{eq:nlp-p-term}
  &0 = p_u &u &\in \Leaves
\end{align}
Here, \eqref{eq:nlp-fix-x}--\eqref{eq:nlp-fix-r}
fix the geometry of the root tree $\Tree_k$
and the locations of all terminal nodes.
Constraints \eqref{eq:nlp-length} and \eqref{eq:nlp-murray}
ensure consistency of $\ell_{uv}$ with $x_u, x_v$ and Murray's law,
respectively, outside $\Tree_k$.
The pressure drop across segment $uv$
and the terminal pressure $p_u$ are given by
\eqref{eq:nlp-deltap} and \eqref{eq:nlp-p-term}, respectively,
where $Q_{uv} = \sum_{vw \in \Arcs} Q_{vw}$
for $v \in \Nodes \setminus (\set{0} \cup \Leaves)$
(Kirchoff's law) and $Q_{uv} = Q\perf / \card\Leaves$ for $v \in \Leaves$
(homogeneous flow distribution).
Moreover, we set $p\term = 0$ without loss of generality.

We use lower bounds $r^- = \SI{0.1}{mm}$,
the radius of vessels entering the microcirculatory network \cite{schreiner1993computer},
and $\ell^- = \SI{0.2}{mm}$ to satisfy the conditions for Poiseuille flow
to hold also for the smallest vessels.
The upper bounds are $\ell^+ = 2 \max_{a \in \Arcs} \ell_a\CCO$
and $r^+ = 2 \max_{a \in \Arcs} r_a\CCO$,
where $\ell_a\CCO, r_a\CCO$ refer to the initial CCO-generated tree.
If the length of a non-terminal segment becomes smaller than its diameter we delete it.
We then replace this degenerate segment with its branch segment, which may create a trifurcation.

\begin{figure}
  \centering
  \subfloat[Standard CCO-driven growth]
    {\image[trim=100 30 25 40,clip=true,width=0.486\textwidth]
    {./Figures/Benchmark/Rectangle_CCO_N_6000} \label{fig:GeoOpt_benchmark_comparison1}
}\hfill
  \subfloat[CCO + geometry optimization via NLP]
  {\image[trim=100 30 25 40,clip=true,width=0.486\textwidth]
    {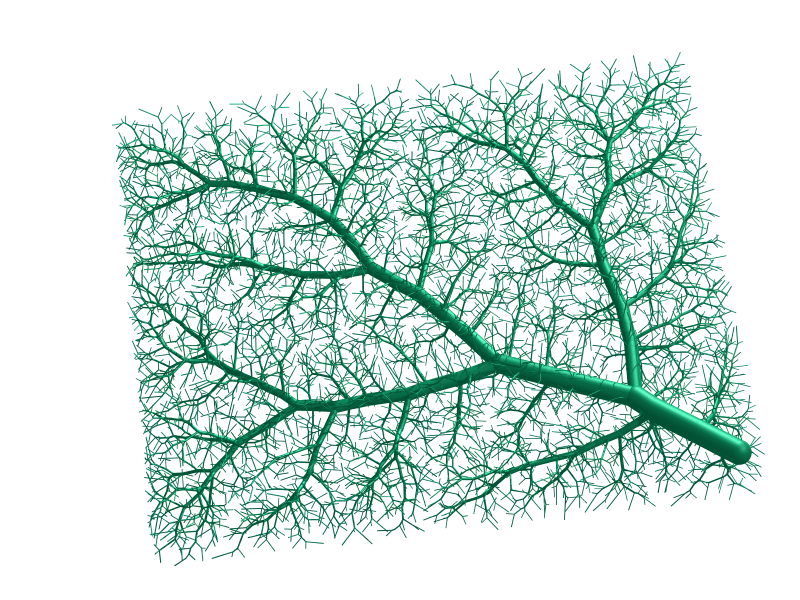} \label{fig:GeoOpt_benchmark_comparison2}
}\\
  \subfloat[Overlaid geometries
    (red = standard CCO, green = CCO + geometry optimization)]
  {\image[trim=30 10 0 10,clip=true,width=0.985\textwidth]
    {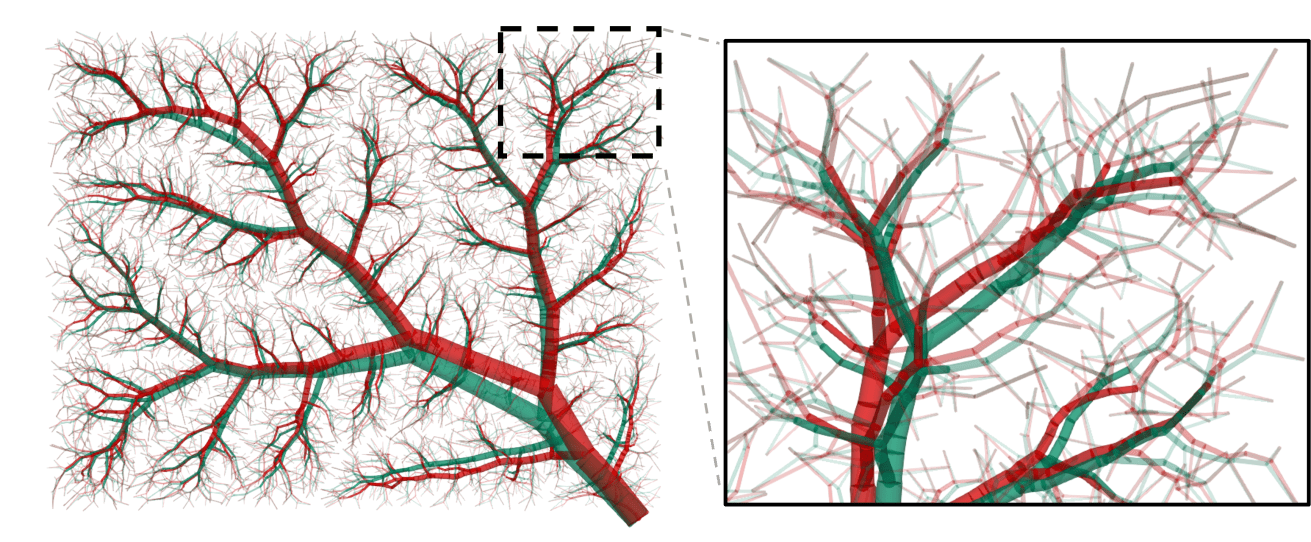} \label{fig:GeoOpt_benchmark_comparison3}
}
  \caption{Comparison of complete tree structures (with $N\term = 6{,}000$)}
  \label{fig:GeoOpt_benchmark_comparison}
\end{figure}

We use our benchmark problem due to Karch et al.\ \cite{karch2000staged}
with $N\term = 6{,}000$ terminal segments
to assess the effect of geometry optimization via the NLP described above.
To this end,
we first compare visualizations of the complete tree structure
generated via standard CCO in \cref{fig:GeoOpt_benchmark_comparison1}
and geometrically optimized afterwards
by solving the NLP in \cref{fig:GeoOpt_benchmark_comparison2}.
We overlay the geometries of both trees in \cref{fig:GeoOpt_benchmark_comparison3}
and observe that, although at this stage the tree topology remains the same,
the two methods lead to significant differences in tree geometry.
As a result of the NLP, the total volume of the tree is reduced by $4.1\%$ compared to the standard CCO tree.
Furthermore, 566 trifurcations are created during the optimization.

As a measure of the symmetry between branches,
the \emph{branching ratio} of node $u$ is defined as \cite{schreiner1993computer}
\begin{equation}
  \delta_u =
  \frac{\min\defset{r_{uv}}{uv \in \Arcs}}{\max\defset{r_{uv}}{uv \in \Arcs}}
  \quad\forall u \in \Nodes \setminus\Leaves.
  \label{eq:bifurcation_ratio}
\end{equation}
To show the impact of repeated geometry optimizations
on the overall branching asymmetry,
we compute the branching ratios \eqref{eq:bifurcation_ratio} over all generations.

\textbf{Remark 1:} To classify the hierarchy throughout the tree,
each segment is assigned to a generation according to the Strahler ordering method \cite{jiang1994diameter}.
The ordering starts from the leaf nodes, which are initially assigned to the order~1.
At each branching, the parent node is assigned the maximum order of its children.
If the children belong to the same order, the parent is assigned the order of its children plus 1. 
For each generation the Strahler order is applied contrariwise, starting with the root segment at generation~1.

\begin{figure}
  \centering
  \image[trim=0 15 0 0,clip=true,width=0.5\textwidth]
  {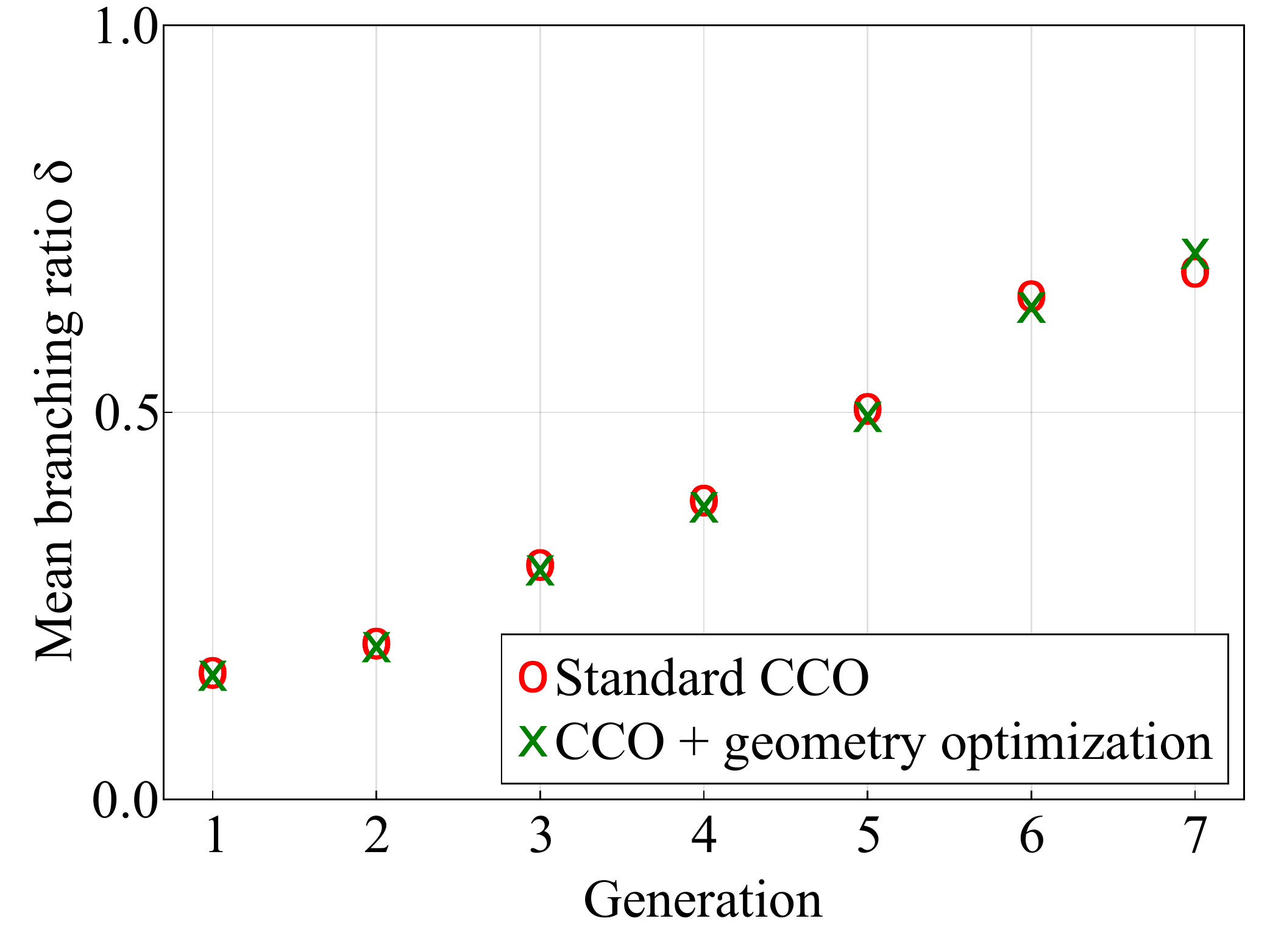}
  \caption{Comparison of branching ratios between
    CCO-generated tree and geometrically optimized tree}
  \label{fig::CCOvsgeo_bifurcation_ratios}
\end{figure}

\cref{fig::CCOvsgeo_bifurcation_ratios} plots the branching ratios for the first seven generations
for the geometrically optimized tree and the standard CCO generated tree.
We observe that optimizing the \emph{global} geometry
improves the branching asymmetry over the generations 2 to 6.
We note that for higher generations, branching ratios of both trees become more symmetric.
This is consistent with observations in corrosion casts \cite{debbaut2014analyzing}, where smaller vessels
also tend to bifurcate more symmetrically.

\begin{figure}[h]
  \centering
  \image[trim=0 5 0 5,clip=true,width=\textwidth]
  {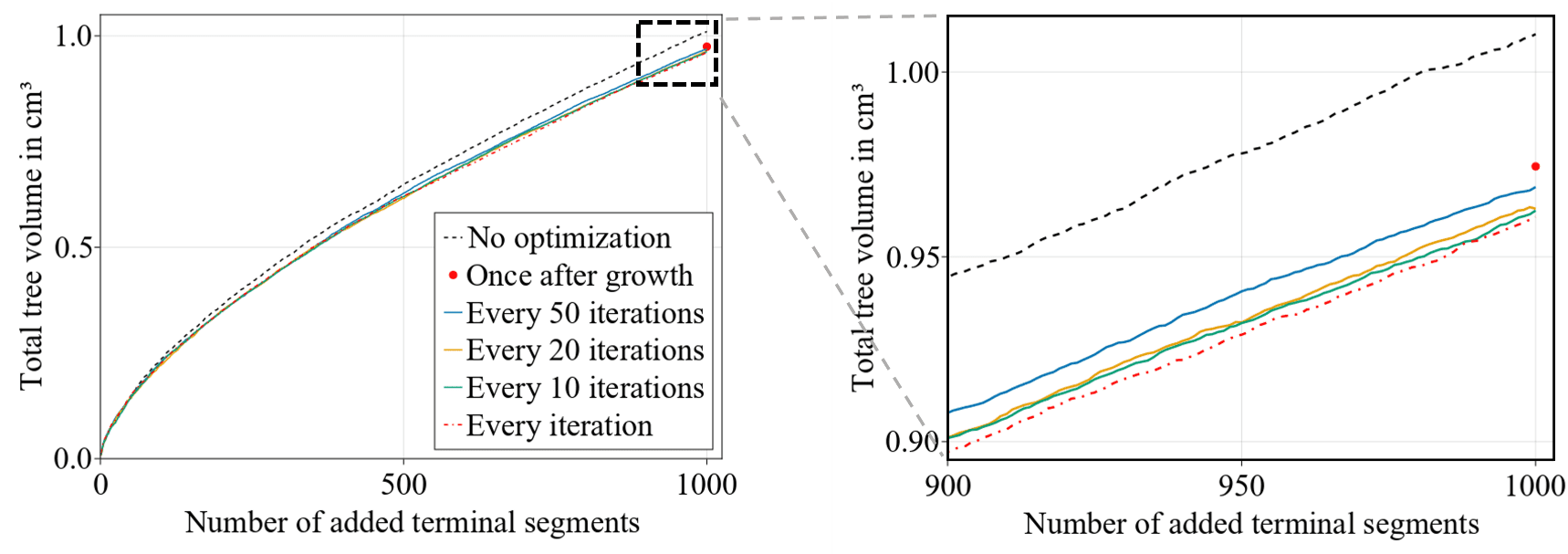}
  \caption{Influence of number of geometry optimizations
    during growth on final tree volume ($N\term = 1{,}000$)}
  \label{fig:GeoOpt_benchmark_growth}
\end{figure}

Optimizing the global geometry each time
after a node is added is computationally expensive.
To reduce the associated computational cost, our idea is to run this optimization after several nodes are added.
To determine an appropriate rule that balances accuracy and computational cost,
we conduct a sensitivity study for the current benchmark problem with $N\term = 1{,}000$.
Based on the results of this study shown in \cref{fig:GeoOpt_benchmark_growth},
we find that carrying out geometry optimization after $N\geo = 20$ new nodes is an appropriate compromise for sparser trees,
which we will increase step-wise during growth to a maximum of 500 for the densest trees (more than 20,000 nodes).
We will apply this rule in all computations in the remainder of this paper.

\subsubsection{Topology optimization}\label{topology_optimization}

We have seen that optimizing the global geometry
reduces the total volume of the vascular tree
and improves its asymmetric branching pattern.
However, the locations of nodes still depend primarily
on the sampling of the terminal points in the CCO algorithm.
To reduce the associated bias,
we propose an additional topology optimization step for an intermediate
tree structure with fewer total segments.
We utilize the property of CCO
that initial samples are not changed significantly during growth
by continuing the growth from this intermediate near-optimal vascular structure.

We optimize the topology by exchanging pairs of proximal points
from one parent segment to another
and then optimizing the global geometry using the NLP model.
This is similar to the local search for the best connection in the standard CCO algorithm,
with the key property of also allowing the swapping of entire sub-trees.
Our topology optimization approach is discrete,
and the total number of possible topologies for a binary branching tree
with $n$ nodes is given by the Catalan number,
\begin{equation}
  C_n = \frac{1}{n+1} \binom{2n}{n}.
\end{equation}
For only $N\term = 500$ segments this still involves
50,000 possible swaps per iteration.
To reduce this number, we delete infeasible swaps that create a cycle
(an ancestor node is connected to the current node)
and swaps where the initial new segment length
is at least two times as large as the current segment length. During tests,
we observed that these swaps almost never lead to improved topologies.
Since the root subtree $\Tree_k$ and the leave locations are kept fixed,
this restricts the search to local topology changes.
The number of possible swaps per iteration then drops to around 7,500.

\begin{figure}[h]
  \centering
  \image[trim=0 10 0 10, clip=true,width=0.5\textwidth]
  {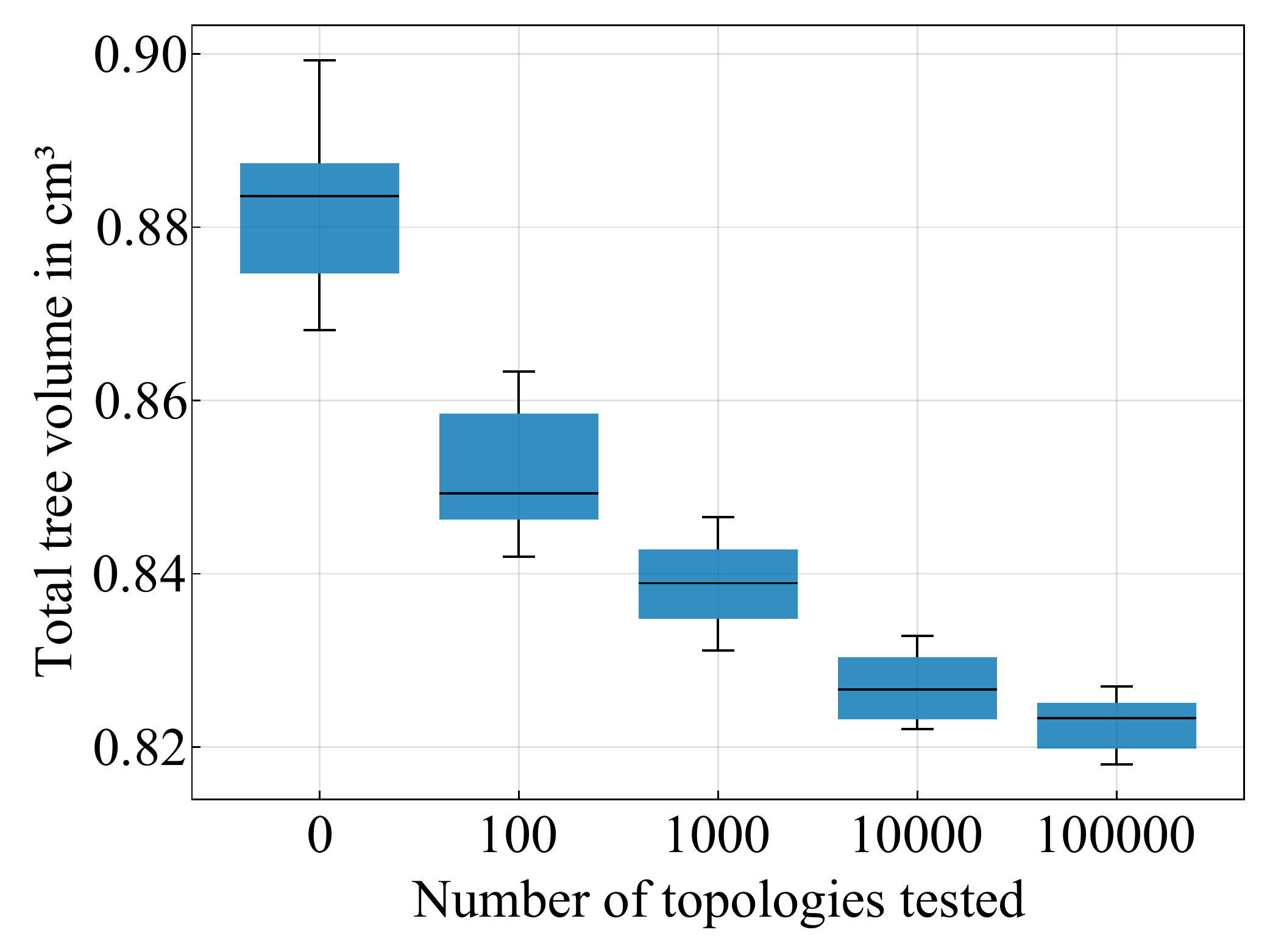}
  \caption{Box plot  of the total volume of 10 trees generated with different seeds
           obtained for different numbers of tested topologies
           during discrete topology optimization ($N\term = 500$)}
  \label{fig:SA_convergence}
\end{figure}

This number is still too large to search the entire possible solution space.
We therefore deploy simulated annealing (SA)
\cite{kirkpatrick1983optimization},
a metaheuristic approach, to search the discrete solution space.
Instead of accepting a new topology only when it yields a smaller volume
than the current one, SA accepts worse topologies with a probability of
\begin{equation}
  p = \exp\left(-\frac{\Delta f_\Tree}{T}\right),
\end{equation}
where $\Delta f_\Tree = f_\Tree^j-f_\Tree^i$ is the change in cost
associated with going from topology $i$ to topology $j$.
$T$ is the SA temperature, which is ``large'' initially and is then ``cooled down'' after each iteration.
This means that SA can ``climb out'' of local minima and search a wider solution space.
\cref{fig:SA_convergence} shows the total volume of 10 different trees
during topology optimization with SA in a box plot, illustrating the effectiveness of the approach.
We observe that not only the topology optimization significantly reduces the total volume,
but also the variance between different trees is reduced.
This indicates that the different random seeds converge to nearly identical tree structures.

Since we use CCO to obtain the initial tree topology,
the initial temperature $T_0$ does not need to be chosen too large,
which significantly reduces computation time.

\subsubsection{Combining geometry and topology optimization}\label{sec:algorithm}

To complete our new optimization framework, we combine geometry and topology optimization.
We specify the perfusion volume to be filled, the number of terminal segments $N\term$
and the inital root subtree of depth $k$ (or proximal point of root for $k = 0$).
After initialization of the problem we use CCO to grow the tree until it has 500 terminal segments.
We optimize the topology of this initial tree, as described in \ref{topology_optimization}.
From this near-optimal tree we restart CCO until $N\term$ segments are added.
After each $N\geo$ iteration steps we optimize the global tree geometry by solving the NLP.
The structure of our framework is shown in \cref{alg:pseudo_code}.

\begin{algorithm}
  \caption{New optimization framework}\label{alg:pseudo_code}
  \begin{algorithmic}[1]
    \For {$i = \card{\Nodes_k}, \ldots, N\term$}
    \State Generate new terminal point $x\term$
    \State Determine best branching location $x$
    on best connection segment $j$ with proximal node $u$
    \If{$\norm{x - x_u} > \ell^-$}
    \State Connect $x\term$ to segment $j$ at position $x$
    \Else
    \State Connect $x\term$ to node $u$ at position $x_u$
    \EndIf
    \If{$i == 500$}
    \State Optimize tree topology using SA approach
    \ElsIf{$i > 500$ and $i \bmod N\geo == 0$}
    \State Optimize global tree geometry by solving NLP
    \State Heuristically increase $N\geo$ based on density of current tree
    \EndIf
    \EndFor
    \State Optimize global tree geometry by solving NLP
    \State Replace degenerate segments with their branch segments
  \end{algorithmic}
\end{algorithm}

Our new optimization framework together with the CCO algorithm was implemented in the programming language \emph{Julia} \cite{bezanson2017julia}.
The NLP is solved by an interior point method using the solver \emph{Ipopt}
\cite{wachter2006implementation} and the linear solver \emph{Mumps}\cite{Amestoy_et_al:2001}.
All computations were done on a desktop computer
with $32$ GB of random-access memory (RAM)
and an Intel Core i9-9900k @5Ghz with $16$ processing threads.

To measure the computation cost of each component of our framework,
we measured the computing times for three different cases.
The first two cases include the Benchmark problem with $N\term = 500$ and $N\term = 6{,}000$ respectively,
and the third is the generation of a portal vein, described in the next chapter, with $N\term = 24{,}000$.
We formally divide our framework in CCO-driven growth,
geometry optimization during growth,
and topology optimization on the reduced tree ($N\term = 500$).
The results are shown in \cref{tab:computing_times}.
It becomes clear that (except for $N\term = 24{,}000$)
the topology optimization using the SA approach is the
most expensive part of the framework,
even though we are limiting it to only $500$ terminal segments.
In contrast, optimizing the global geometry
during growth is efficient even for the portal vein problem.
It takes \SI{45}{s} to solve the NLP for 24,000 terminal segments.

\begin{table}[hb]
  \centering
  \caption{Computing times of new
    optimization framework for three different cases}\label{tab:computing_times}
  \begin{tabular}{l*5r}
    \toprule
    & \multicolumn{2}{c}{Benchmark} & \multicolumn{1}{c}{Portal vein} \\
    & ($N\term = 500$) & \quad($N\term = 6{,}000$) & \quad($N\term = 24{,}000$) \\
    \midrule
    CCO-driven growth & \SI{10}{s} & \SI{565}{s} & \SI{8640}{s} \\
    Geometry optimization & \SI{15}{s} & \SI{285}{s} & \SI{2930}{s} \\
    Topology optimization & \SI{4820}{s} & \SI{4970}{s} & \SI{6126}{s} \\
    \bottomrule
  \end{tabular}
\end{table}

\begin{figure}
  \centering
  \subfloat[Geometry optimization only (via series of NLPs)]
  {\image[trim=100 30 25 40,clip=true,width=0.495\textwidth]
    {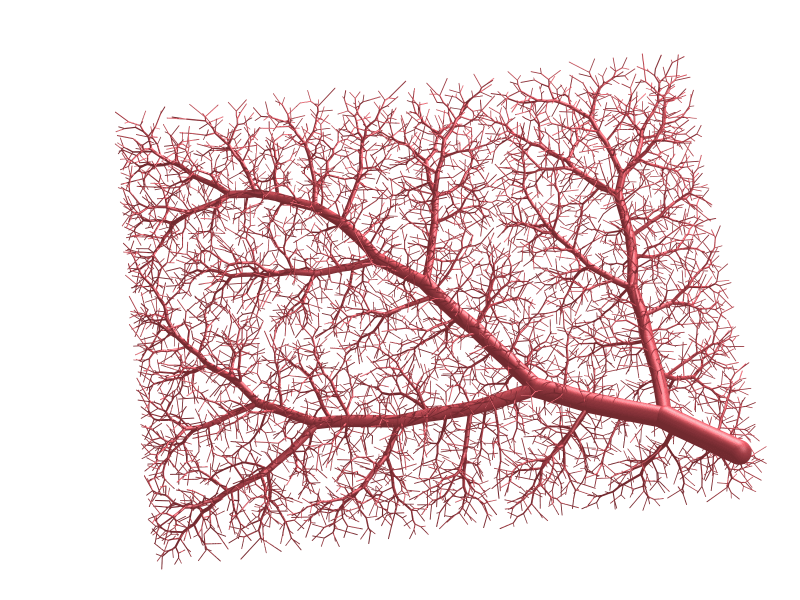}}\hfill
  \subfloat[Geometry and topology optimization (via series of NLPs + discrete topology testing)]
  {\image[trim=100 30 25 40,clip=true,width=0.495\textwidth]
    {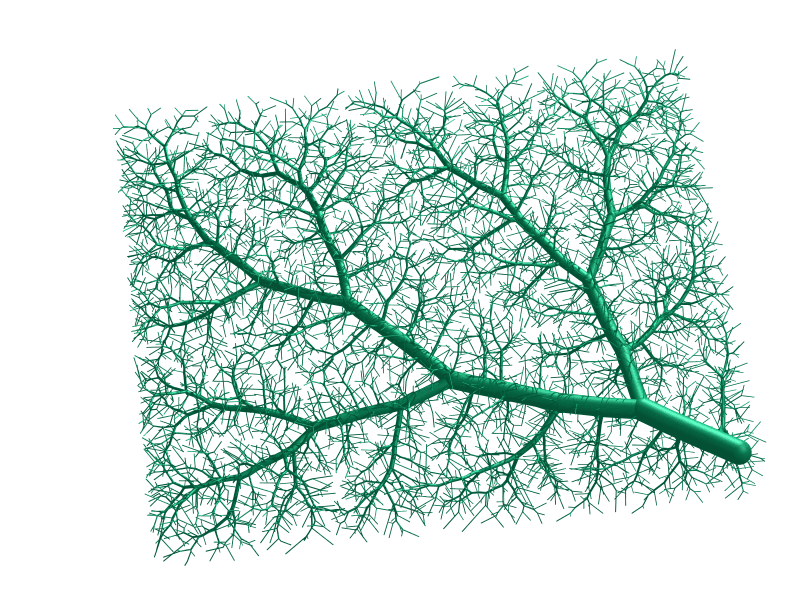}}\\
  \subfloat[Overlaid geometries
    (red = geometry optimization only, green = geometry and topology optimization)]
  {\image[trim=30 10 0 10,clip=true,width=0.985\textwidth]
    {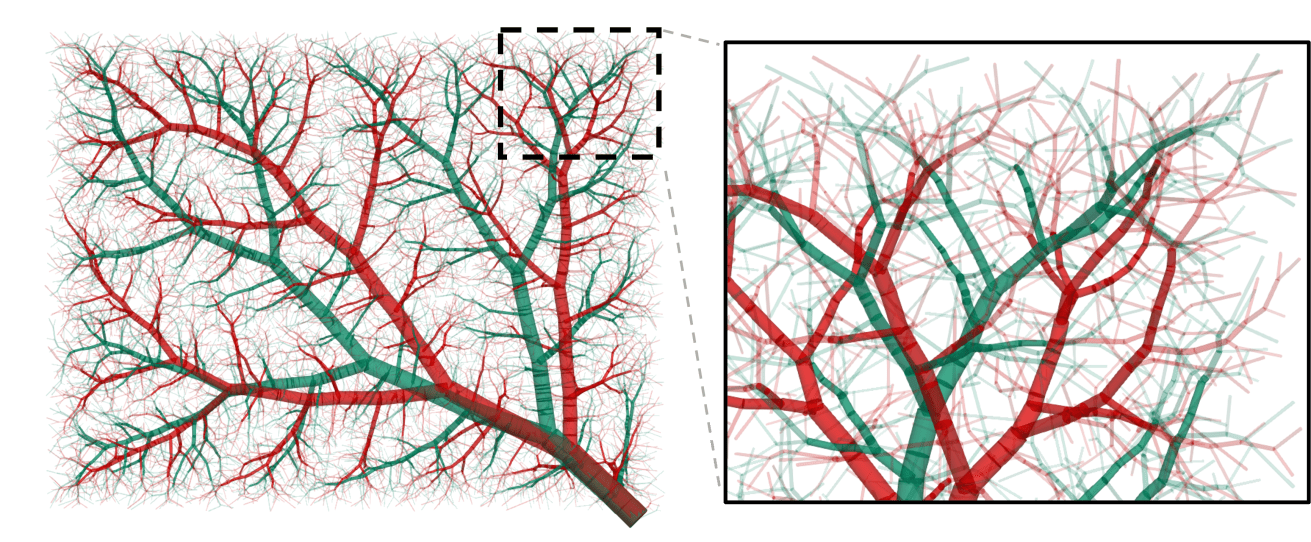} \label{TopoOpt_benchmark_comparison3}}
  \caption{Vascular trees before
    and after topology optimization ($N\term = 6{,}000$)}
  \label{fig:TopoOpt_benchmark_comparison}
\end{figure}

Using our current benchmark example,
\cref{fig:TopoOpt_benchmark_comparison} enables a visual comparison of the complete vascular tree
that is geometrically optimized via solving the NLP and the complete vascular tree
that is geometrically and topologically optimized.
We observe  that the geometrically and topologically optimized tree differs significantly
from the tree that is only geometrically optimized.

\begin{figure}
  \centering
  \subfloat[Seed 1 (geometrically optimized)]
  {\image[trim=100 30 25 40,clip=true,width=0.495\textwidth]
    {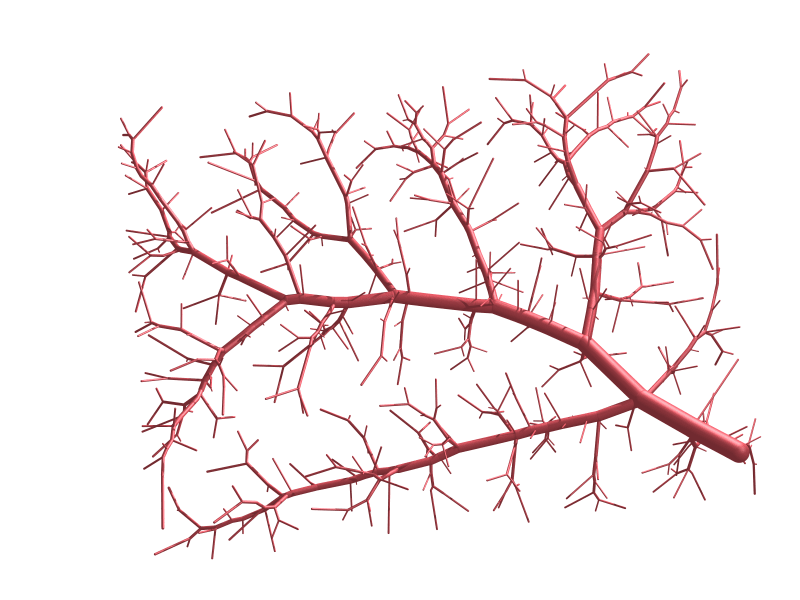}%
    \label{fig:CCO_benchmark_random_results_2_1}}\hfill
  \subfloat[Seed 1 (geometrically + topologically optimized)]
  {\image[trim=100 30 25 40,clip=true,width=0.495\textwidth]
    {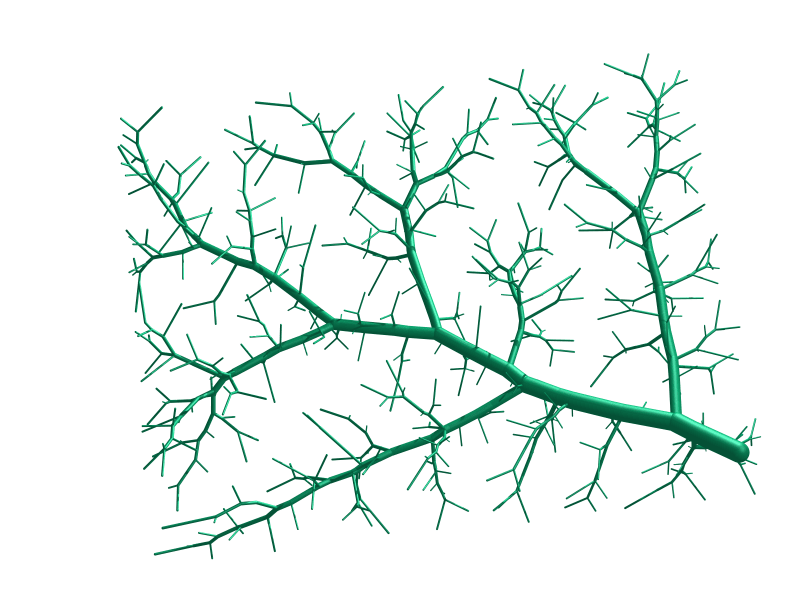}%
    \label{fig:CCO_benchmark_random_results_2_1opt}}\\
  \subfloat[Seed 2 (geometrically optimized)]
  {\image[trim=100 30 25 40,clip=true,width=0.495\textwidth]
    {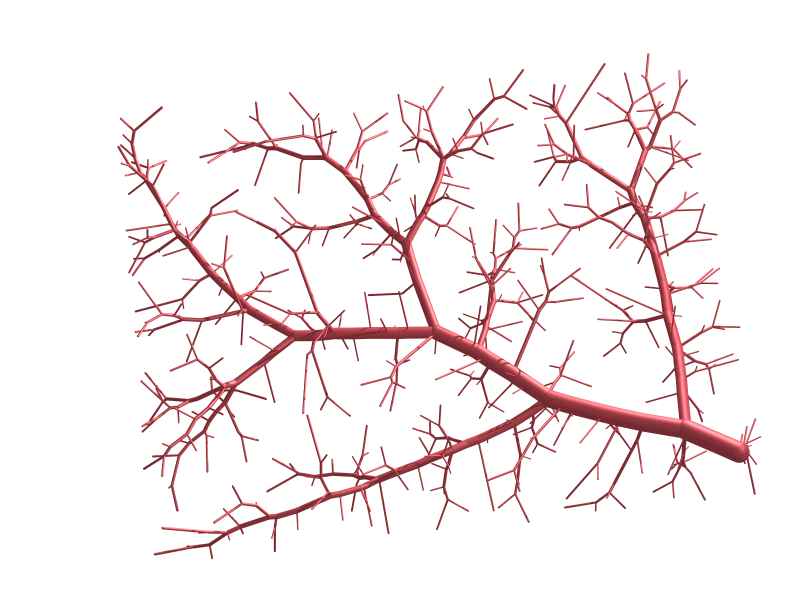}%
    \label{fig:CCO_benchmark_random_results_2_2}}\hfill
  \subfloat[Seed 2 (geometrically + topologically optimized)]
  {\image[trim=100 30 25 40,clip=true,width=0.495\textwidth]
    {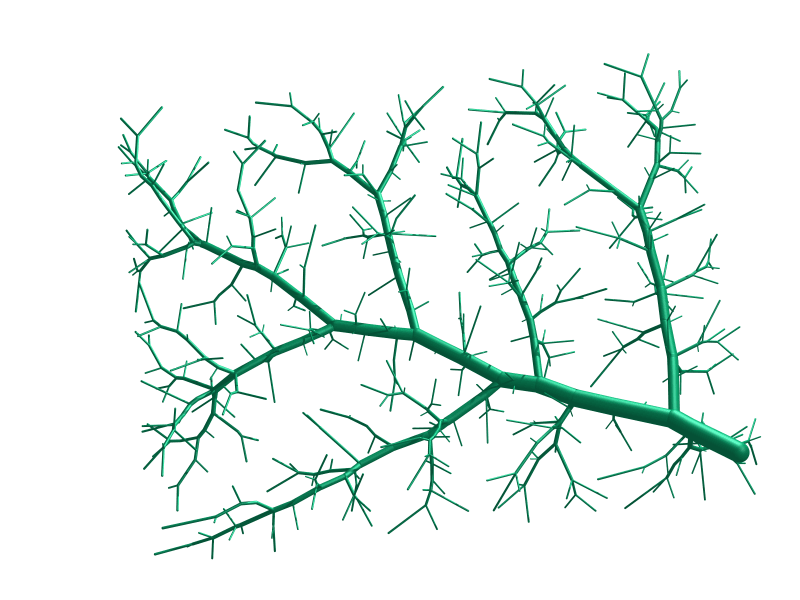}%
    \label{fig:CCO_benchmark_random_results_2_2opt}}\\
  \subfloat[Seed 3 (geometrically optimized)]
  {\image[trim=100 30 25 40,clip=true,width=0.495\textwidth]
    {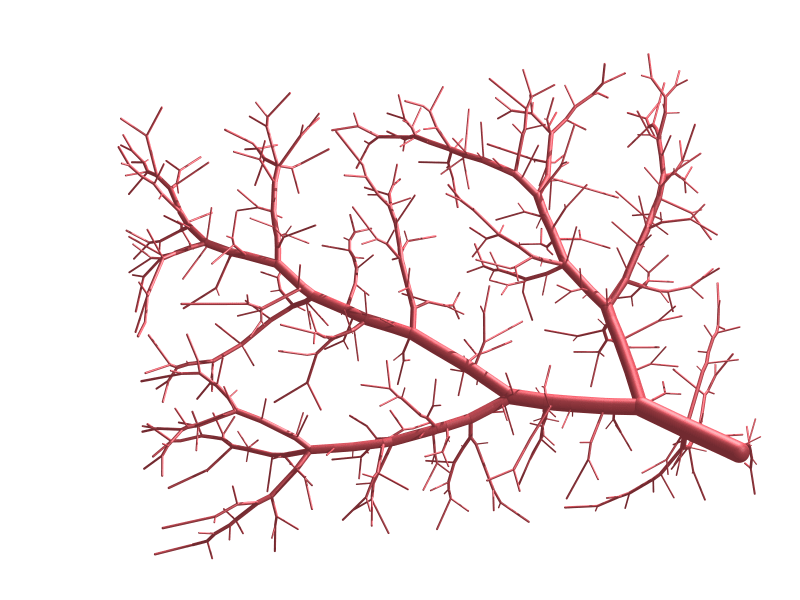}%
    \label{fig:CCO_benchmark_random_results_2_3}}\hfill
  \subfloat[Seed 3 (geometrically + topologically optimized)]
  {\image[trim=100 30 25 40,clip=true,width=0.495\textwidth]
    {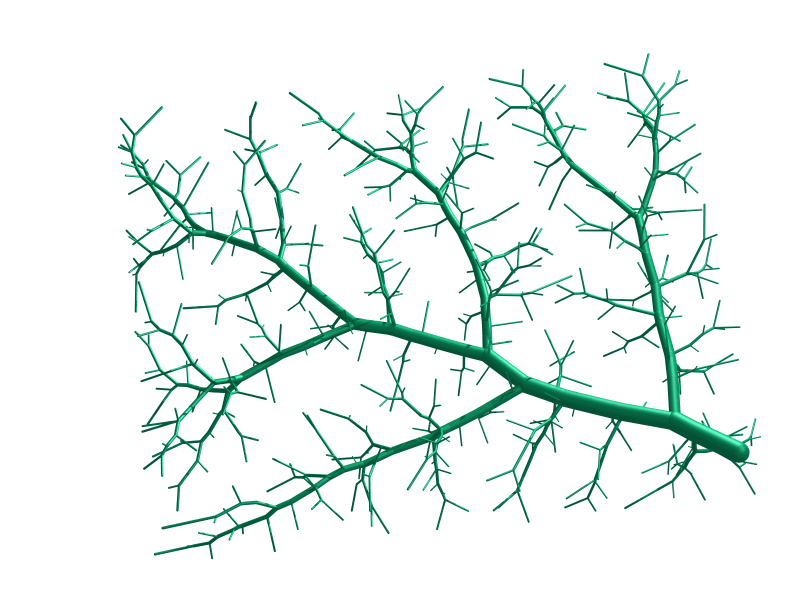}%
    \label{fig:CCO_benchmark_random_results_2_3opt}}\\
  \caption{Trees generated with different random seeds
    before and after topology optimization ($N\term = 500$)}
  \label{fig:CCO_benchmark_random_results_2}
\end{figure}

To better illustrate the importance of topology optimization,
we consider the three geometrically optimized trees with $N\term=500$
that are shown in the left column of \cref{fig:CCO_benchmark_random_results_2}.
They are juxtaposed to the corresponding versions
after having applied the topology optimization.

\begin{figure}
  \centering
  \subfloat[Seeds (geometrically optimized)]
  {\image[trim=70 20 65 30,clip=true,width=0.49\textwidth]
    {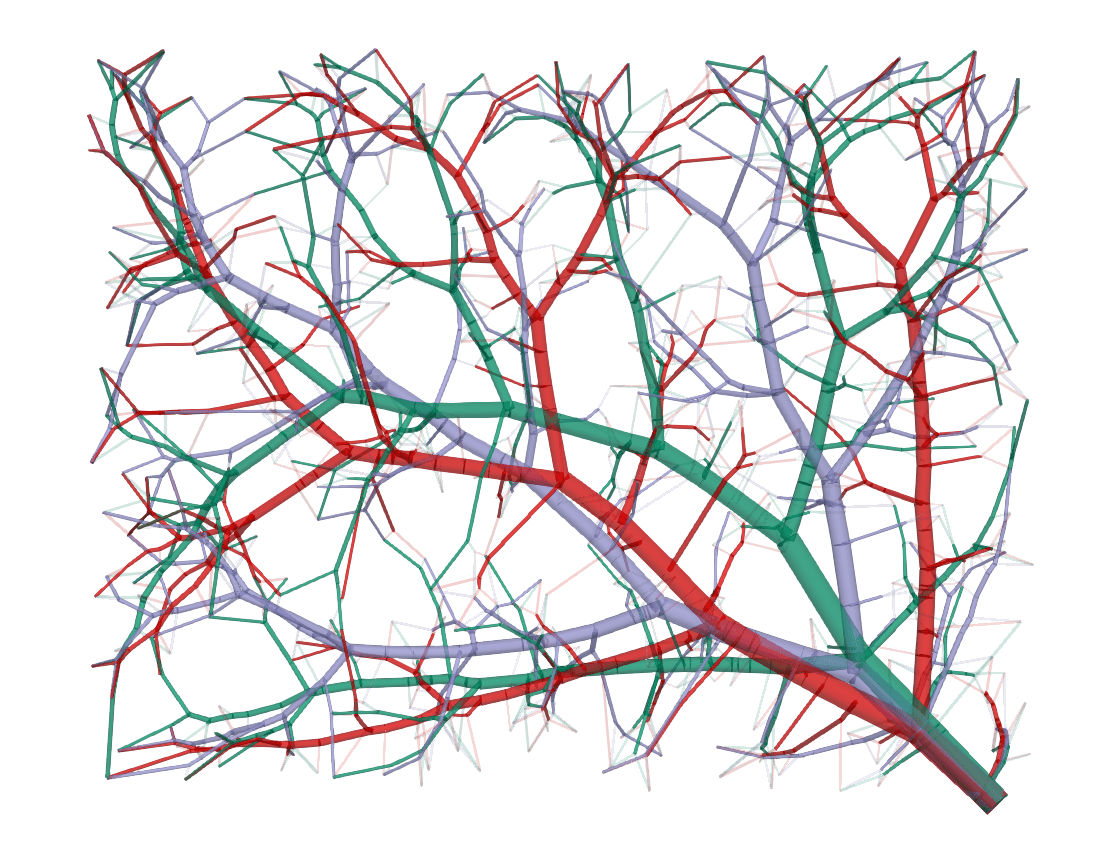}%
    \label{fig:Geo_vs_topo_overlay1}}\hfill
  \subfloat[Seeds (geometrically + topologically optimized)]
  {\image[trim=70 20 65 30,clip=true,width=0.49\textwidth]
    {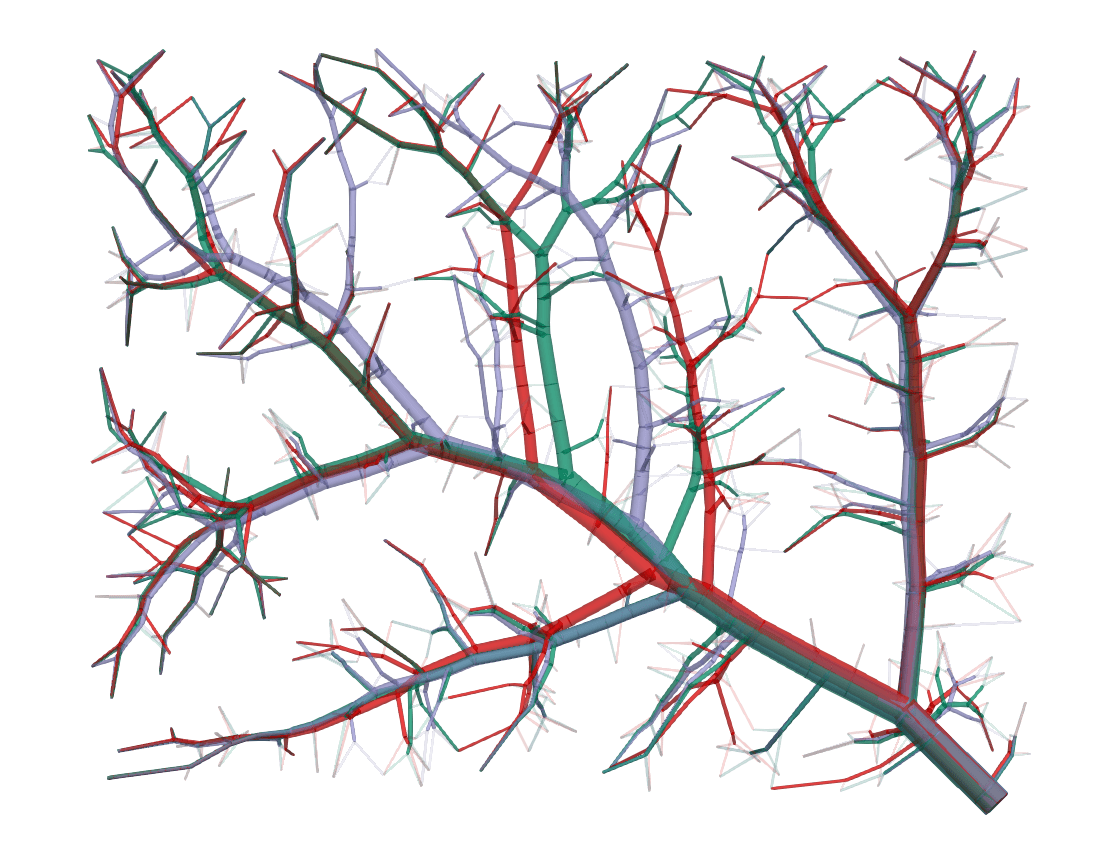}%
    \label{fig:Geo_vs_topo_overlay2}}
  \caption{Overlaid geometries of different random seeds
    before and after topology optimization
    ($N\term = 500$; green = Seed 1, red = Seed 2, purple = Seed 3)}
  \label{fig:Geo_vs_topo_overlay}
\end{figure}

For the current benchmark, topology optimization further reduces the total volume of the tree
by up to $6\%$, resulting in a total volume decrease of up to $11\%$ with respect to the standard CCO-generated trees.
We can also observe in \cref{fig:CCO_benchmark_random_results_2} that all three trees,
although generated with different random seeds, converge towards very similar tree structures.
This convergence is also highlighted in \cref{fig:Geo_vs_topo_overlay},
where we overlaid the different trees before and after topology optimization, respectively.
In particular, we see in all three results a prominent large trunk going from the bottom right corner to the top left corner,
connecting two main branches on the topside and one main branch on the bottom side.

\section{Validation}

\begin{figure}
  \centering
  \image[trim=12 10 10 10,clip=true,width=0.4\textwidth]
  {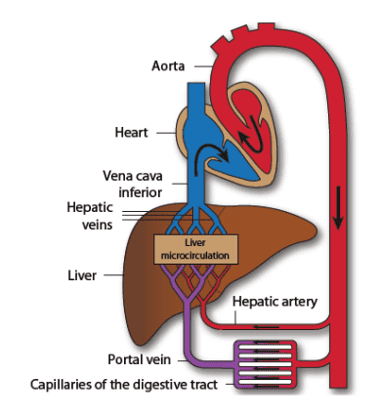}
  \caption{Schematic overview of the liver inside the systemic circulation from Debbaut et al.\ \cite{debbaut2014multi}}
  \label{fig:liver_organism_scale}
\end{figure}

We have developed a framework based on mathematical optimization
that allows us to generate synthetic vascular trees with reproducible topology
and geometry for general non-convex perfusion volumes.
We can now validate the overall approach against real vascular systems.
To this end, we consider the hepatic vascular systems in the human liver.
Blood flow through the liver on the organism scale is shown in \cref{fig:liver_organism_scale}.
In contrast to other organs, the liver has two supplying trees.
The first one is supplied through the \emph{hepatic artery} (HA)
from the heart,
and the second one is supplied through the \emph{portal vein} (PV)
from the digestive tract.
The blood leaves the liver through a single draining tree
into the \emph{hepatic veins} (HV) leading into the vena cava inferior (VCI).

In the scope of this work, we focus on the supplying tree that stems from the portal vein.
We apply our framework for generating a synthetic hepatic tree that we can then assess
via a real hepatic tree experimentally characterized via a detailed vascular corrosion cast of a human liver.

\subsection{Vascular corrosion casting}

As in-vivo medical imaging cannot provide detailed representations of hepatic tree structures,
we resort to ex-vivo vascular corrosion casting, as described in detail by Debbaut et al.\ \cite{debbaut2014analyzing}.
The ex-vivo liver (weight $\approx$\SI{1.9}{kg})
was first connected to a machine perfusion preservation device
from Organ Recovery Systems in Zaventem, Belgium.
During a 24 hour period, the liver was continuously perfused under pressure-control
through the HA at \SI{25}{mmHg}
and the PV at \SI{7}{mmHg}.
The blood left the liver through the HV and VCI.
The perfusion of the liver allows the preservation of the vasculature and parenchyma. Moreover, it keeps the blood vessels open.
The color-dyed casting resin was added to both the HA and PV
simultaneously until a sufficient quantity emerged from the VCI.
Afterwards, inlet and outlet vessels were clamped to avoid resin leakage
during the polymerization step, which took approximately 2 hours.
After two days of a macerating bath, the corrosion cast was ready for imaging.
The liver cast was imaged \emph{in globo} and the resulting image dataset was reconstructed using Octopus software (Ghent University, Gent, Belgium).
The complete casting and micro-CT setup is illustrated in \cite{debbaut2014analyzing}. More detailed information on the vascular corrosion casting and micro-CT scanning can be found in \cite{debbaut2010vascular}.

\begin{figure}
  \centering
  \subfloat[Segmented image of all three vascular trees]
  {\image[trim=70 60 70 30,clip=true,width=0.495\textwidth]
    {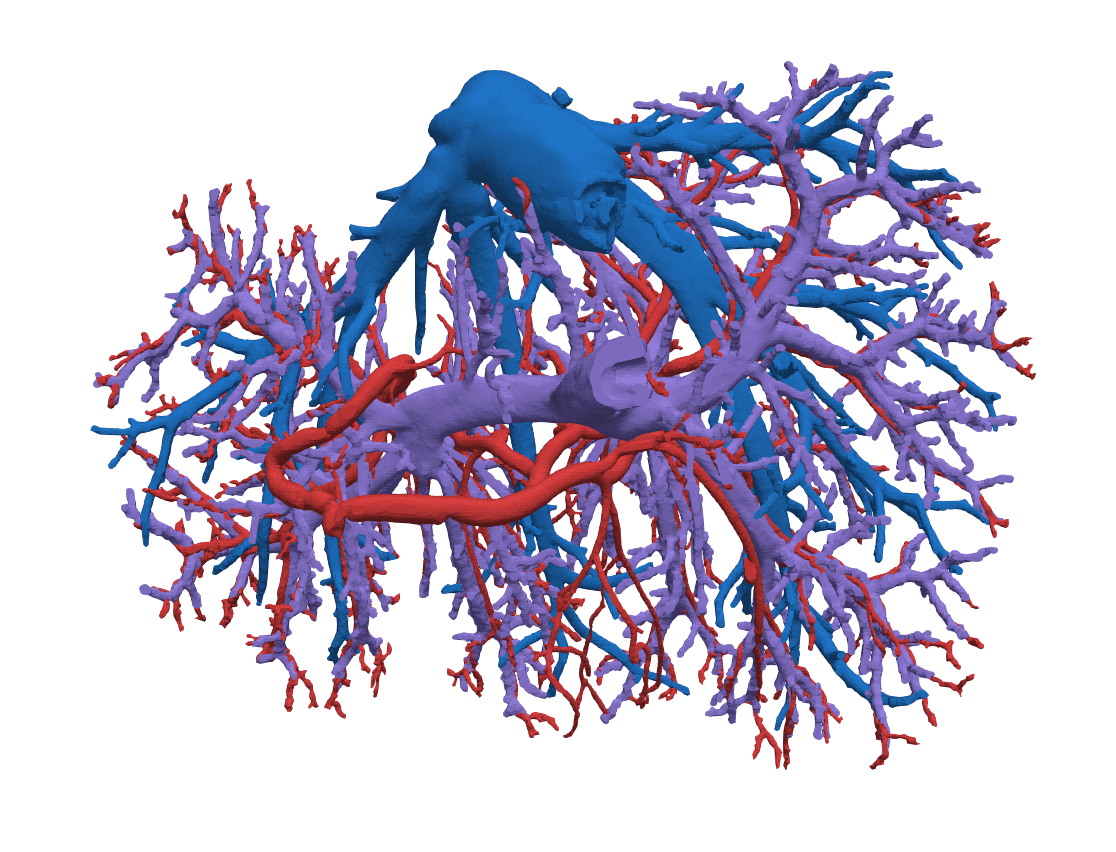}}\hfill
  \subfloat[Segmented image of PV]
  {\image[trim=70 60 70 30,clip=true,width=0.495\textwidth]
    {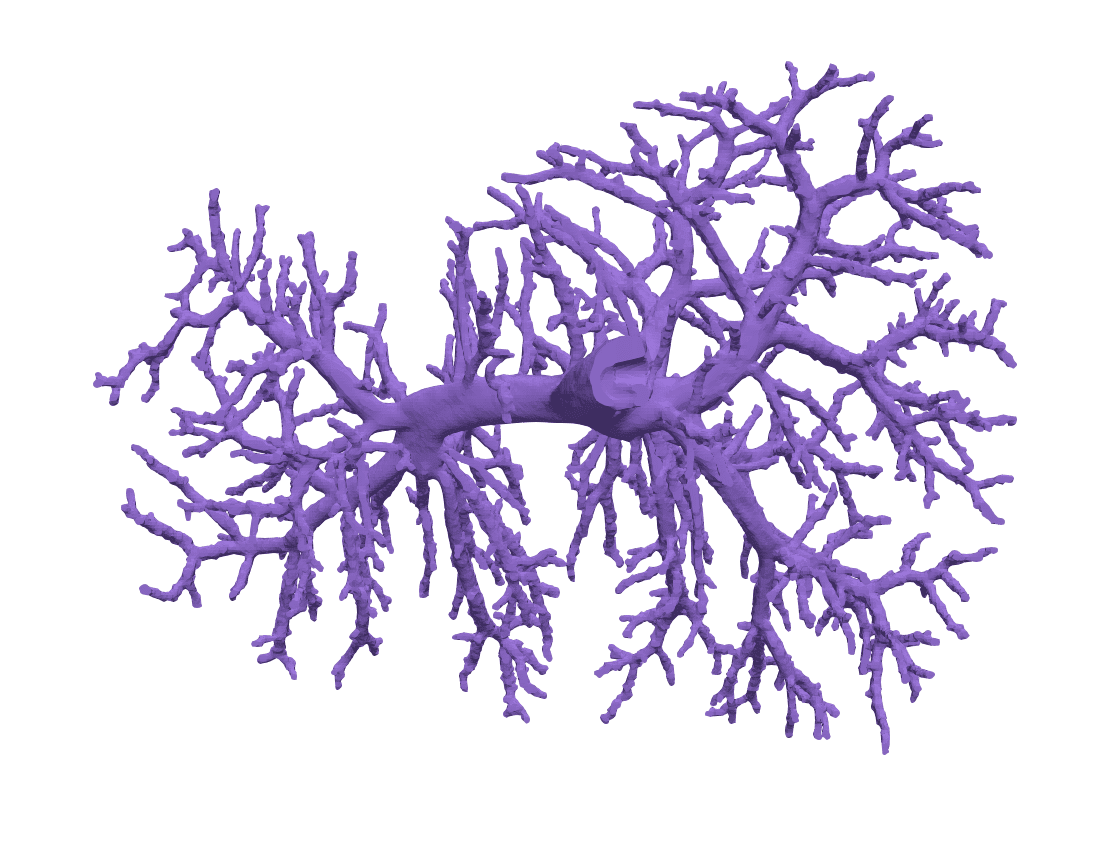}}\\
  \subfloat[Segmented image of HV]
  {\image[trim=70 60 70 30,clip=true,width=0.495\textwidth]
    {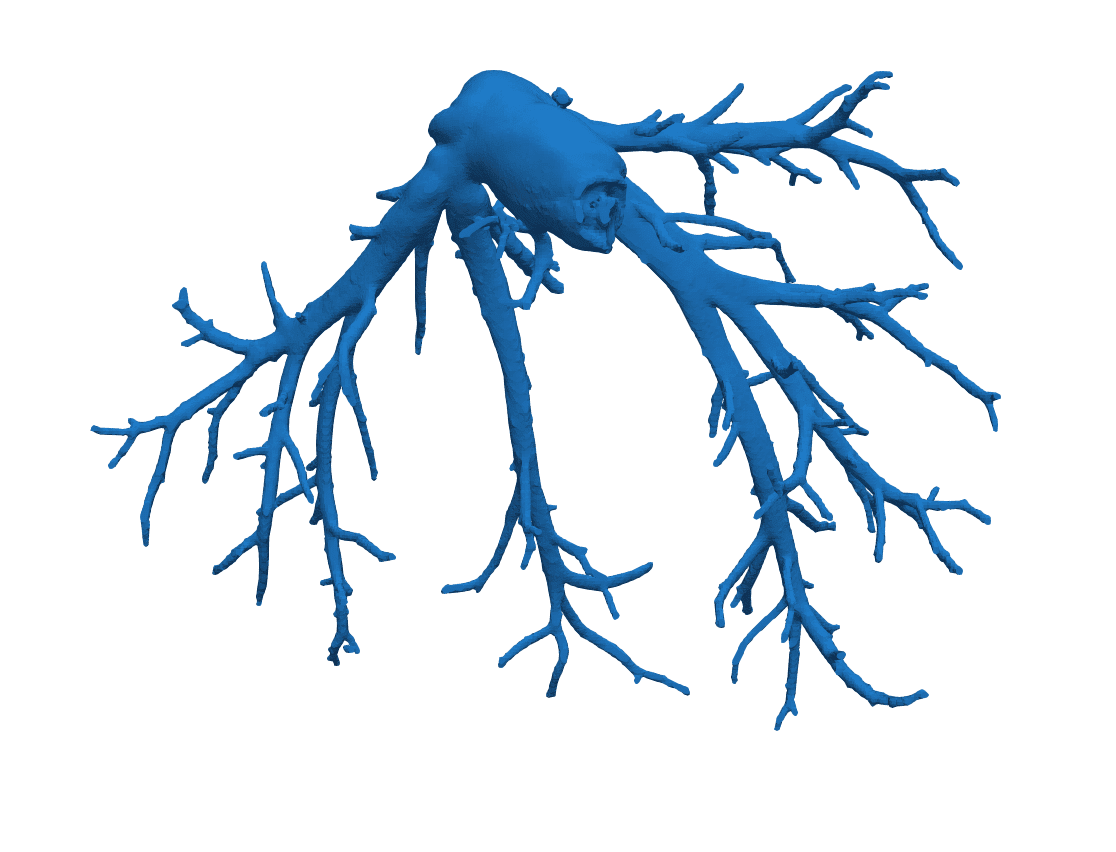}}\hfill
  \subfloat[Segmented image of HA]
  {\image[trim=70 60 70 30,clip=true,width=0.495\textwidth]
    {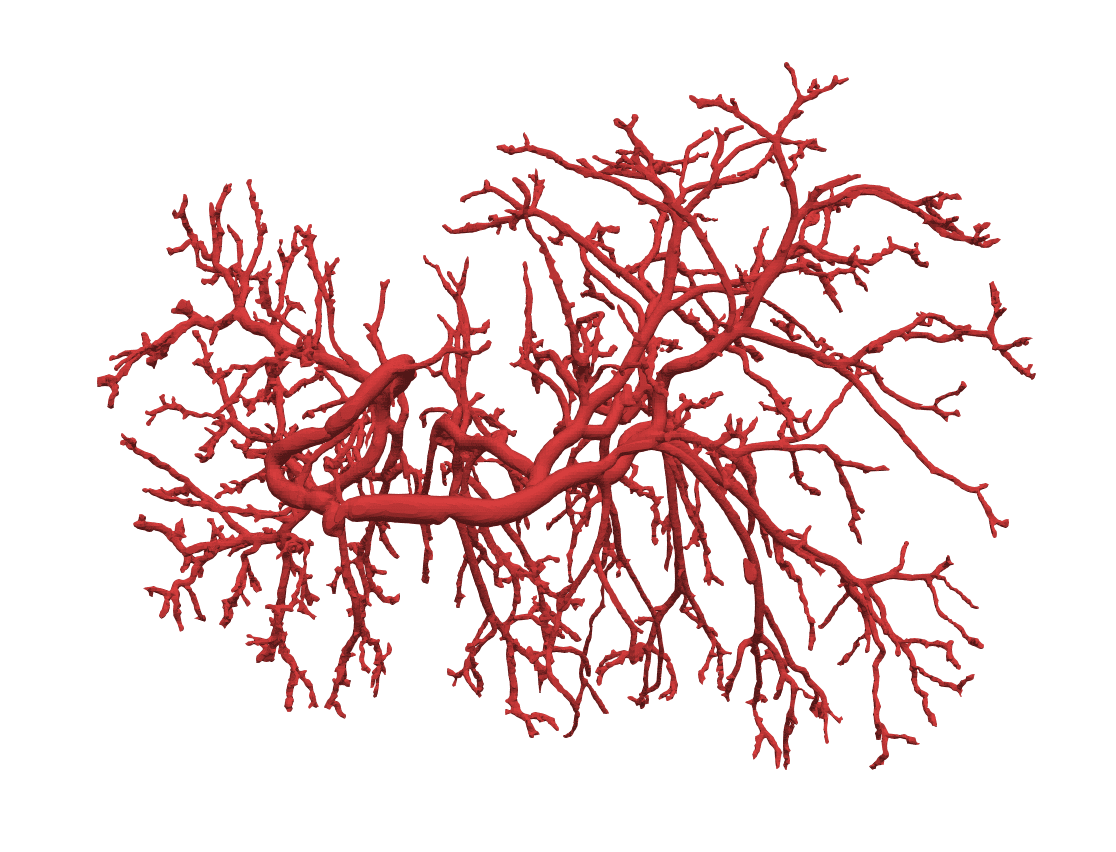}}
  \caption{Representations of all three vascular hepatic trees obtained from imaging of the corrosion cast as obtained in \cite{debbaut2014analyzing}}
  \label{fig:corrosion_cast_results}
\end{figure}

The resulting micro-CT data set was processed and segmented
based on the gray values of the images.
Due to the contrast agent used in the HA resin, the separation
of arterial and venous vessels was facilitated.
The separation of PV and HV trees, however,
was more challenging due to similar gray values and touching vessels, needing manual segmentation.
After the segmentation, a three-dimensional reconstruction of each tree was calculated.
The resulting geometries are shown in \cref{fig:corrosion_cast_results}.

A detailed visual inspection of the tree representations shows that in addition to bifurcations,
all trees also exhibit a number of trifurcations.
We also observe monopodial branches sprouting from parent vessels at angles close to $90^\circ$.
After the first generations, the HA vessels typically run parallel to the PV vessels. This trend continues down to the microcirculation. From the macro- to mesocirculation, mean radii decreased to 0.08 mm at the most distal mesocirculation generation 13 in the sample studied in \cite{debbaut2014analyzing}. At the microcirculation level, blood reaches the functional units of the liver, called hepatic lobules.
This smallest scale of the circulation exhibits entirely different flow characteristics \cite{peeters2017multilevel}
that we cannot describe with our model. Instead, more specific models
as in \cite{ricken2015modeling} would be needed.

\subsection{Comparison and assessment}

The synthetic generation of the PV tree is based
on the perfusion volume of the experimentally investigated tree from Debbaut et al.\ \cite{debbaut2014analyzing}
and the physiological parameters taken from Kretowski et al.\ \cite{kretowski2003physiologically};
see \cref{tab:PV_parameters}.
We generate the vascular tree with $N\term = 24{,}000$ segments,
both with the standard CCO method and with our new framework
as described in \cref{sec:algorithm}.
Our framework takes \SI{4}{hours} \SI{50}{min},
while the standard CCO method takes \SI{2}{hours} \SI{37}{min}; see \cref{tab:computing_times}.

\begin{table}
  \centering
  \caption{Physiological parameters required for the generation of a hepatic vascular tree (portal vein), adapted from
    Kretowski et al.\ \cite{kretowski2003physiologically}}
  \label{tab:PV_parameters}
  \begin{tabular}{*2{l@{\qquad}}l}
    \toprule
    Parameter & Meaning & Value \\
    \midrule
    $V\perf$ & perfusion volume & $\approx$\SI{1500}{cm^3} \\
    $p\perf$ & perfusion pressure & \SI{12}{mmHg} \\
    $p\term$ & terminal pressure & \SI{8}{mmHg} \\
    $N\term$ & number of terminal segments & 6,000 \\
    $Q\perf$ & perfusion flow (at root) & \SI{1000}{ml/min} \\
    $\eta$   & blood viscosity & \SI{3.6}{cP} \\
    $\gamma$ & branching exponent & $3.0$ \\
    $N\con$  & number of connections tested & $30$ \\
    \bottomrule
  \end{tabular}
\end{table}

\begin{figure}
  \centering
  \subfloat[Number of vessels]{\image[width=0.495\textwidth]
    {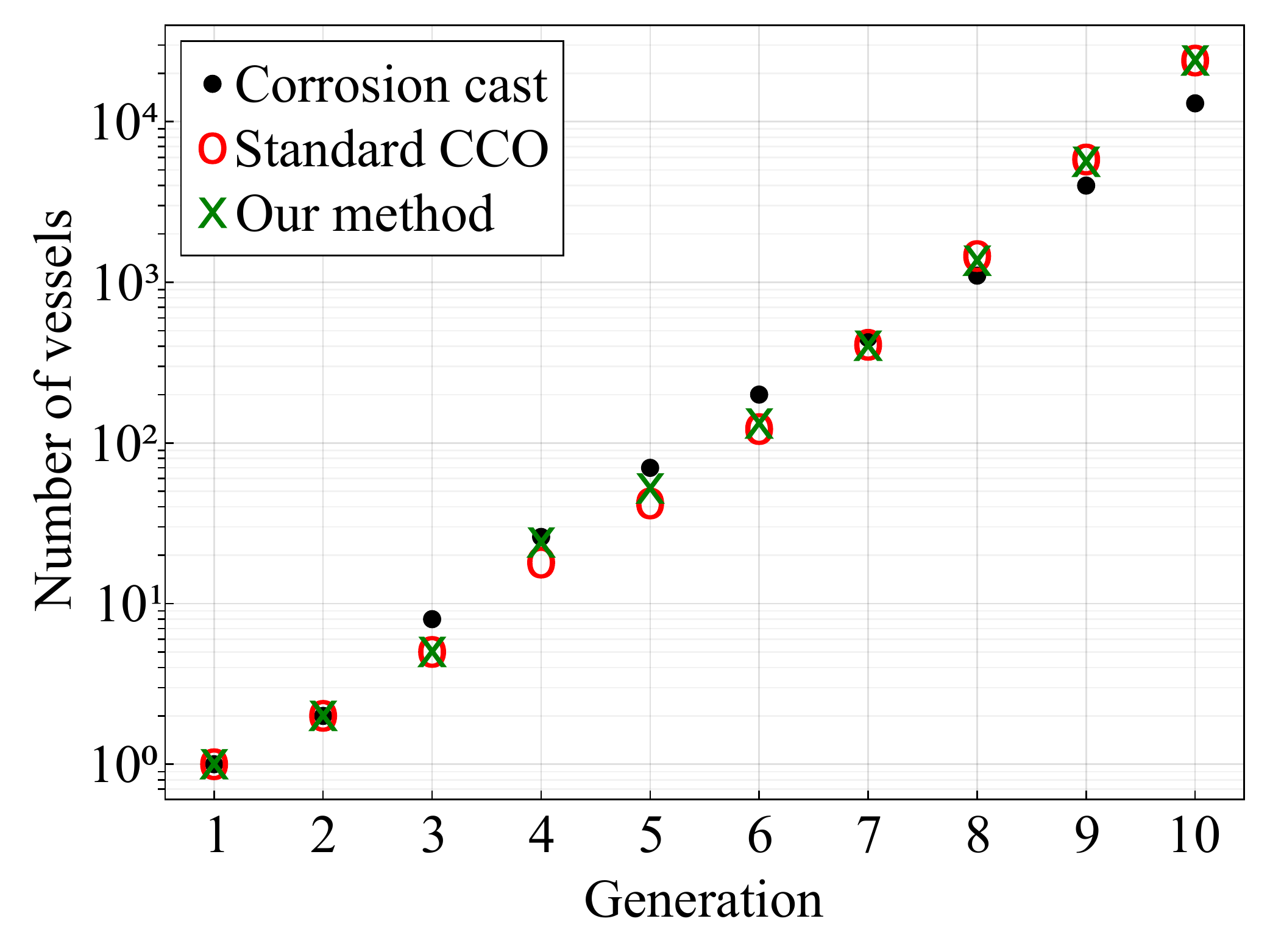}%
    \label{fig:PV_tree_statisticsa}}\hfill
  \subfloat[Mean radii]{\image[width=0.495\textwidth]
    {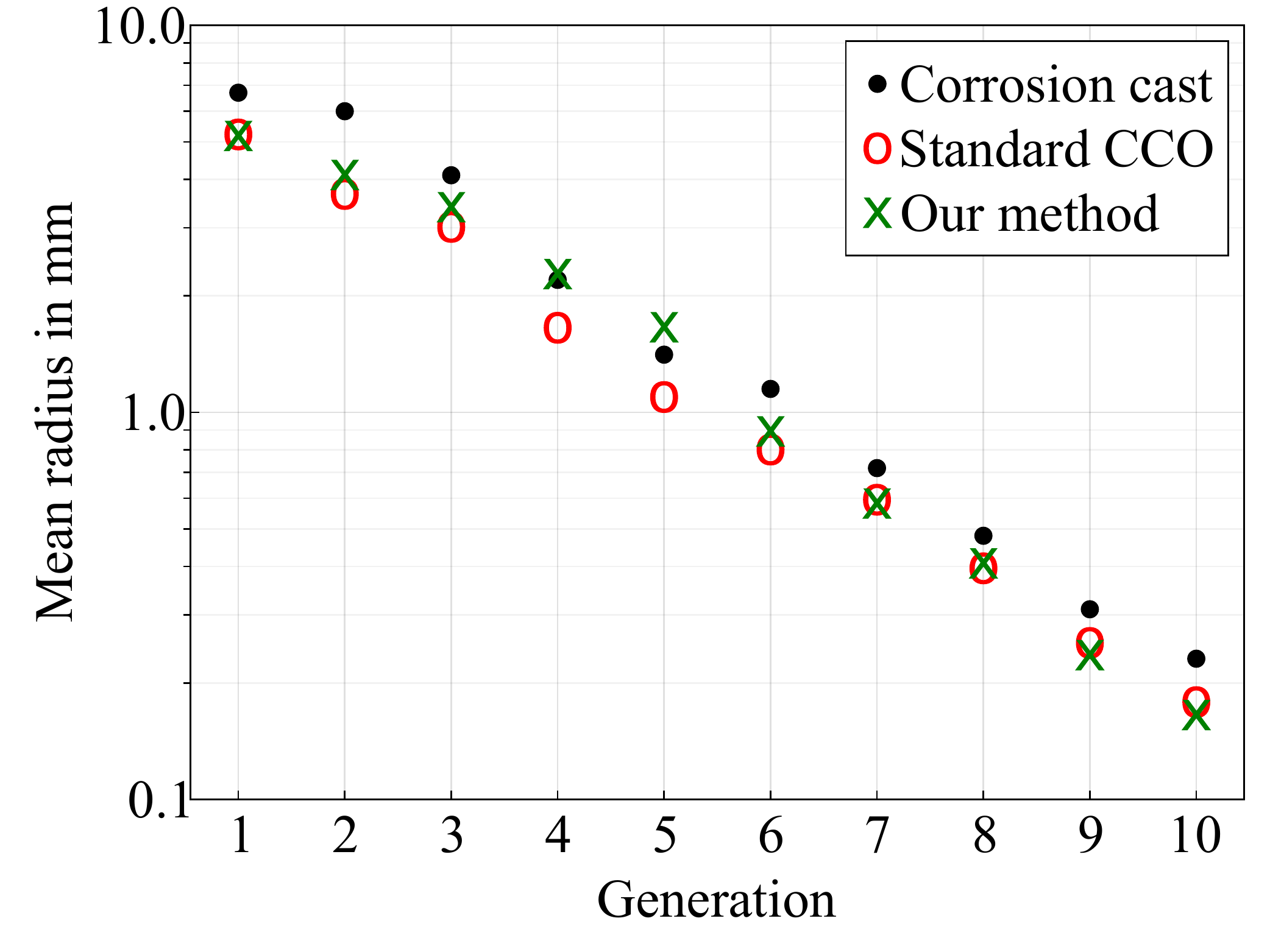}%
    \label{fig:PV_tree_statisticsb}}\\
  \subfloat[Mean branching ratios]{\image[width=0.495\textwidth]
    {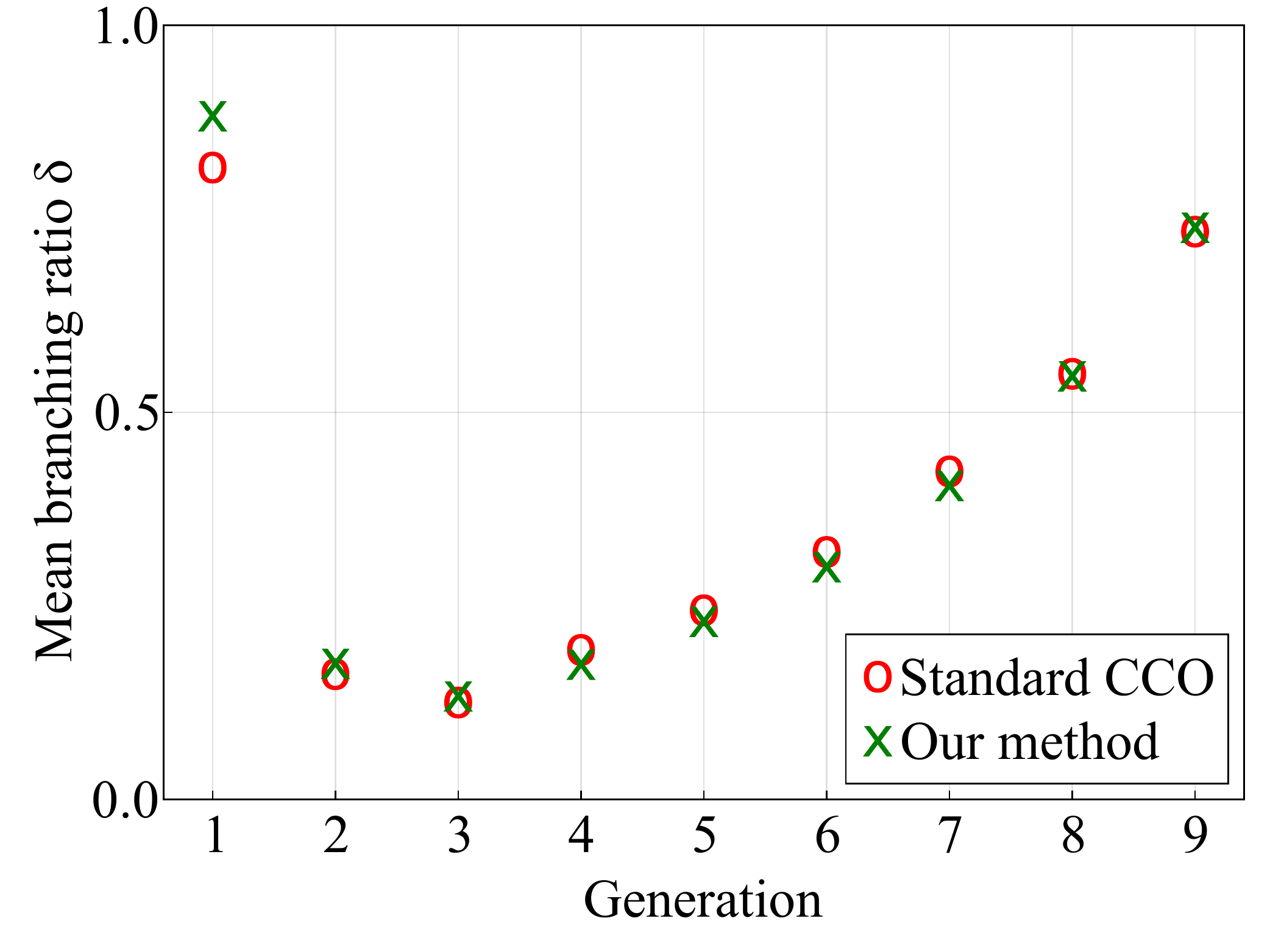}%
    \label{fig:PV_tree_statisticsc}}
  \caption{Key statistics of the portal vein tree:
    our method vs.\ CCO and corrosion cast measurements}
  \label{fig:PV_tree_statistics}
\end{figure}

We start the analysis by a qualitative comparison of the segment parameters averaged over each generation.
In \cref{fig:PV_tree_statistics}, we compare the number of segments and segment radii between standard CCO,
our method based on optimizing the global geometry,
and the reference values calculated by Debbaut et al.\ \cite{debbaut2014analyzing} based on corrosion cast measurements,
for each generation of the hierarchical tree structure.
We can observe in \cref{fig:PV_tree_statisticsa}
that the number of vessels per generation deviates only slightly between all three cases.
Comparing the average radius per generation in \cref{fig:PV_tree_statisticsb},
however, indicates that our method fits the corrosion cast data better
than the standard CCO results for the important lower generations between 1 and 6.
We hypothesize that this improvement is due to the fact
that optimizing the global geometry
shortens the overall segment length of the intermediate generations,
leading to larger radii overall.
In contrast, CCO overestimates the lengths in these generations due to the
limiting view of optimizing the local geometry only,
which leads to smaller radii overall.
For higher generations beyond 7, both methods seem to underestimate the corrosion cast data.
We hypothesize that this observation is due to the fact that the experimental values for the higher generations were interpolated from a small mesoscale sample in the corrosion cast, possibly overestimating the actual mean values.
The improvement in the branching asymmetries in \cref{fig:PV_tree_statisticsc} is also significant, especially for the generations 4 to 6. The high branching ratio for generation 1 signifies that the root segment branches symmetrically into daughter branches with similar radius.
The branching asymmetries for the lower generations increased,
leading to an increase in monopodial branches and an overall higher number of thicker vessels,
which is also visually more comparable to the corrosion cast.
The branchings ratios tend to be larger for the higher generations
and are comparable for both the standard CCO and our method.
This is also backed up by Debbaut \cite{debbaut2014analyzing} in the corrosion cast of a smaller mesoscale sample.

\begin{figure}
  \centering
  \subfloat[Corrosion cast]
  {\image[trim=0 0 0 0,clip=true,width=1.0\textwidth]
    {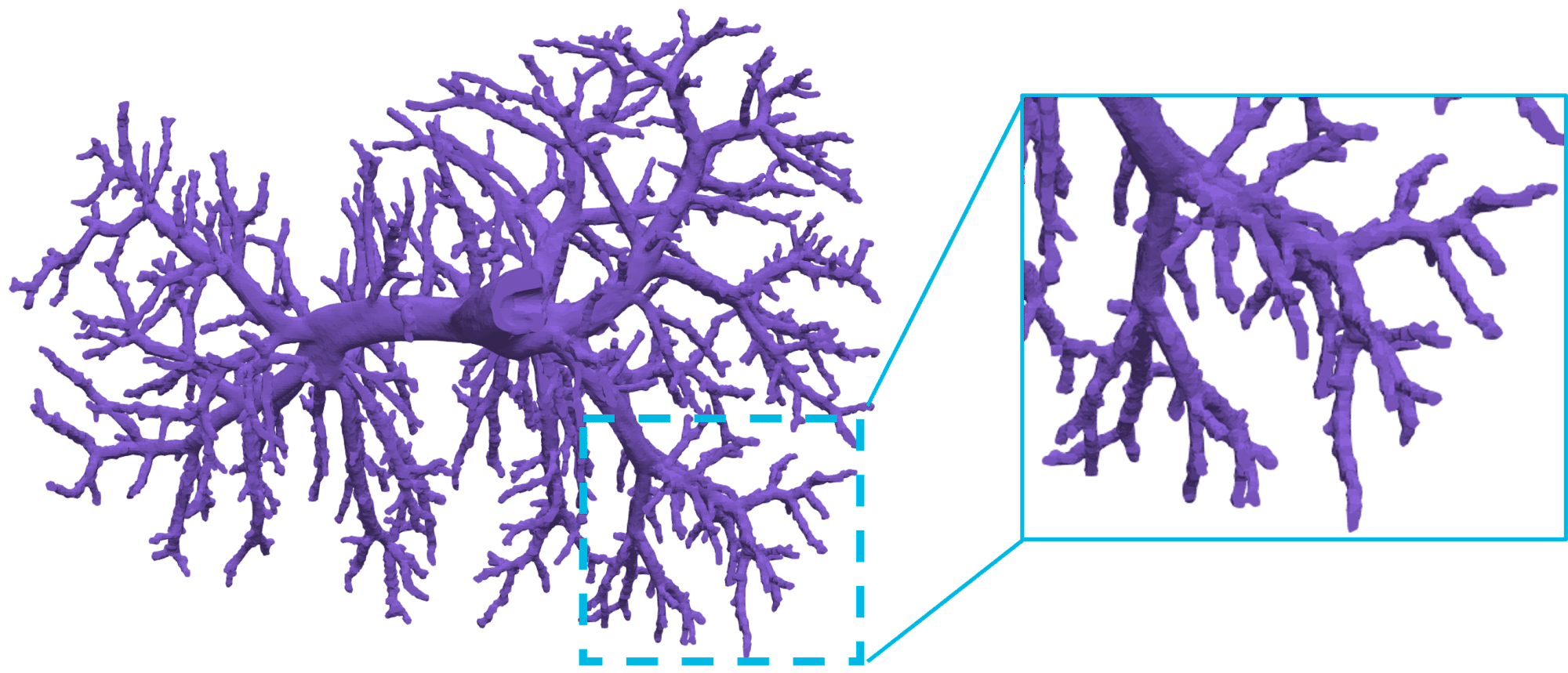}}\\
  \subfloat[Our method]
  {\image[trim=0 0 0 0,clip=true,width=1.0\textwidth]
    {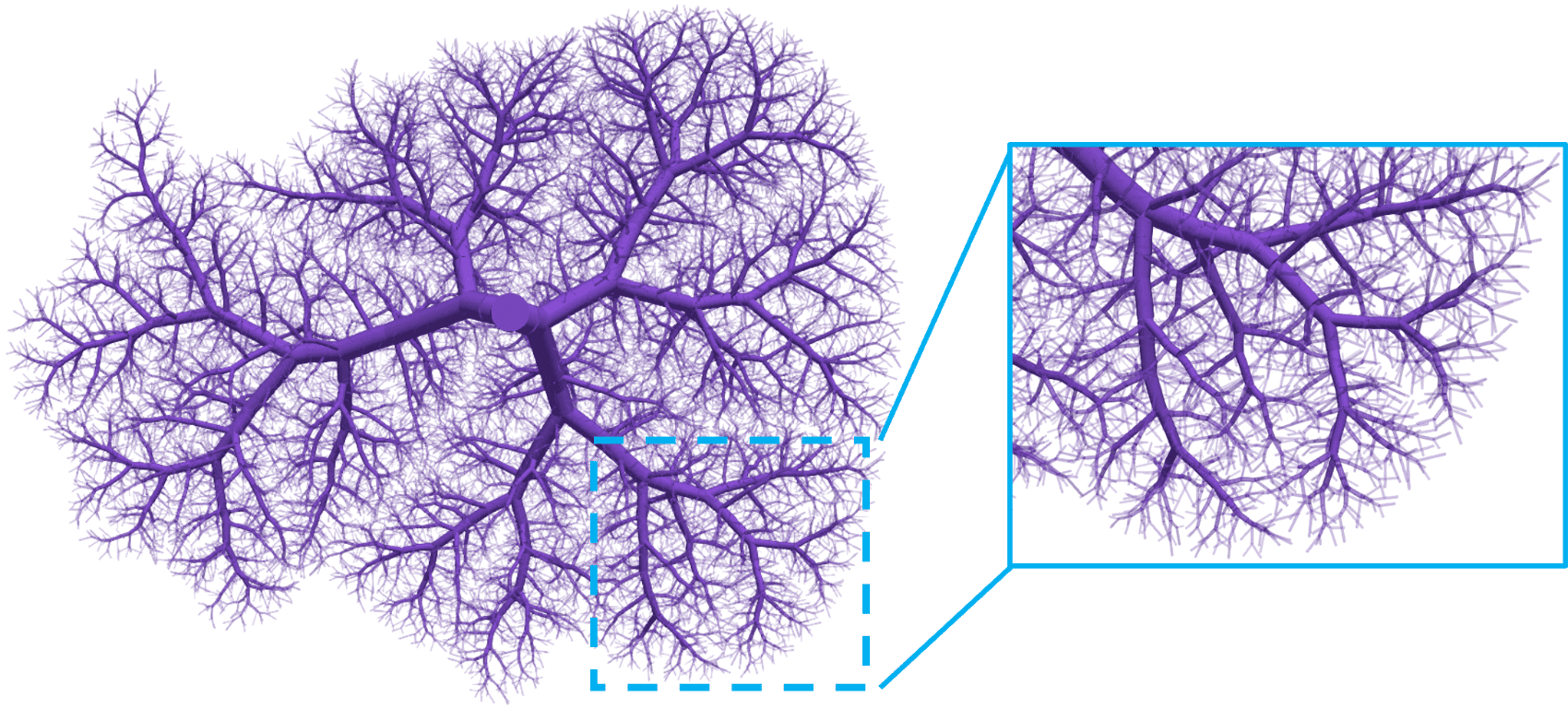}}
  \caption{The complete portal vein tree:
    synthetically generated tree structure vs.\ corrosion cast data}
  \label{fig:comparison_corrosion_cast}
\end{figure}

Finally, \cref{fig:comparison_corrosion_cast} shows the synthetically generated tree structure of the portal vein
and the corrosion cast data below each other.
In addition to bifurcations, the corrosion cast data exhibits 34 trifurcations.
In our method, the tree exhibits 41 trifurcations over the first seven generations, whereas in standard CCO trifurcations are impossible by design.
Furthermore, the number of monopodial branches increased from $341$ to $521$ from the standard CCO tree to our tree.
Lastly, the visual comparison of the synthetic tree structure
based on optimizing the global geometry with the corrosion cast data shows good agreement,
especially for the early generations.
In particular, in both trees, the root vessels split horizontally (with respect to the depicted view)
and have seven major arteries (generations 2--3) splitting from there.
Zooming in to the bottom right corner, we observe that both trees show highly similar branching patterns.
We also see, however, that in other areas, there are pronounced differences.
For instance, the bottom center of the synthetic tree is supplied uniformly via a larger vessel that diagonally stretches downwards,
whereas the corresponding area in the corrosion cast is nearly empty.
We hypothesize that the reason for this difference is due to the missing hepatic vein
and hepatic artery that are not taken into account in the synthetic model,
but are of course present in the corrosion cast, see \cref{fig:corrosion_cast_results}.
This region is also close to where the gall bladder is typically situated.

\section{Summary, conclusions and outlook}
The core assumption behind the synthetic generation of vascular trees is
that their physiological formation is governed by optimality principles to reduce the overall metabolic demand.
Current synthetic tree generation methods
such as constrained constructive optimization (CCO) are capable
of reproducing qualitative measures of their real counterparts,
but fail to achieve comparable branching patterns.
Furthermore, due to dependence on random sequences,
methods such as CCO cannot guarantee
reproducibility of their results,
making a quantitative comparison and validation nearly impossible.
We showed that these drawbacks also stem from the fact that standard methods
such as CCO are based only on optimizing the local tree structure.

In this paper, we developed a new powerful framework for generating synthetic vascular trees to mitigate the above limitations.
The fundamental basis of our framework is the search for a  minimum
in both the tree's global geometry and global topology.
In contrast to standard methods,
we split this search into a distinct  geometry optimization and a  topology optimization.
This allows us to formulate the geometry optimization as a nonlinear optimization problem (NLP).
Unlike other methods, this permits efficient solution algorithms such as the interior point method,
vastly improving the overall computation time.
We combine CCO with a subtree-swapping procedure for the topology optimization
to search between different topologies iteratively.
In each iteration, we optimize the geometry of the new topology by solving the NLP.
We use a metaheuristic algorithm, similar to simulated annealing,
to either accept or reject a new topology.
Finally, we combine these steps into a single algorithmic approach.

Our new algorithm is capable of generating synthetic trees with up to 11 generations.
As input, we only need the (non-convex) volume that is perfused and the root segment's entry point.
The resulting trees showed improved branching patterns while reducing the metabolic cost by up to $11\%$.
Furthermore, results are reproducible, and the influence of random seeds on the global structure is significantly reduced.
This allowed us to directly compare a synthetic hepatic tree against the portal vein of a liver corrosion cast.
Our comparison showed similar branching patterns and comparable geometric locations of both the segments and branchings.
In areas where the influence of the geometry of the hepatic veins is not strong, these similarities reach down to the fifth generation.
Also, the number of trifurcations and monopodial branches formed during growth
is close to that of the real hepatic tree characterized by the corrosion cast data.

The direct comparison with the corrosion cast data
also showed some limitations of the current framework
that we would like to address in future work.
Formally, we can categorize these into model-related,
application (liver) related, and method-related.
On the model part, we made significant assumptions,
namely for the blood viscosity and the cost function.
The blood viscosity should take the Fåhræus--Lindqvist effect into account.
The cost function only considers the total volume as the minimization goal.
In addition, further factors such as the transport cost of blood
should be critically evaluated as additional optimization goals.
For the liver application, results clearly showed that the hepatic vein tree
has a significant influence on the geometry of the portal vein tree.
This will be similar in other organs
with clearly defined inflow and outflow trees.
As such, our framework should be extended to allow
the generation of both trees in a coupled manner.
All these extensions of the framework will certainly increase
the overall computational complexity.
This means that the method's efficiency must be further improved.
Currently, using the NLP model for geometry optimization
is both robust and efficient,
and CCO combined with the heuristic subtree swapping procedure
is a good practical approach for searching the discrete space.
However, a proper mixed-integer nonlinear optimization model (MINLP)
for the topology optimization would be desirable.
Although solving such a rigorous formulation is extremely hard
and would require a substantial mathematical research effort,
it might ultimately produce better topologies,
and it could even provide optimality certificates for the solutions.

\section{Declarations}
\textbf{Funding and/or Conflicts of interests/Competing interests:}
The results presented in this paper were achieved as part of the ERC Starting Grant project ``ImageToSim'' that has received funding from the European Research Council (ERC) under the European Union’s Horizon 2020 research and innovation programme (Grant agreement No.~759001). The authors gratefully acknowledge this support.

\newpage
\bibliographystyle{ieeetr}
\bibliography{references}

\end{document}